\documentclass[12pt]{spieman}  % 12pt font required by SPIE;
\usepackage{amsmath,amsfonts,amssymb}
\usepackage{graphicx}
\usepackage{setspace}
\usepackage{tocloft}
\usepackage{lineno}

\usepackage[colorlinks=true, allcolors=blue]{hyperref}
\usepackage{multirow}
\usepackage{aasmacros}

\makeatletter
\let\c@lofdepth\relax
\let\c@lotdepth\relax
\makeatother
\usepackage{subfigure}

\title{MORFEO control strategy}

\author[a,f,*]{Guido Agapito}
\author[a,f]{Lorenzo Busoni}
\author[a,f]{Cédric Plantet}
\author[a,f]{Giulia Carlà}
\author[a,f]{Alfio Puglisi}
\author[b]{Jean-Pierre Véran}
\author[c,d]{Marco Falotico}
\author[c,d]{Gabriele Umbriaco}
\author[e]{Edoardo Redaelli}
\author[c]{Paolo Ciliegi}
\affil[a]{INAF Osservatorio Astrofisico di Arcetri, Largo E. Fermi 5, 50125 Firenze, Italy}
\affil[b]{Herzberg Astronomy and Astrophysics, National Research Council of Canada, 5071 West Saanich Road,
Victoria, BC V9E 2E7, Canada}
\affil[c]{INAF Osservatorio di Astrofisica e Scienza dello Spazio di Bologna (OAS), via Gobetti 93/3, 40129 Bologna, Italy}
\affil[d]{Alma Mater Studiorum - Università di Bologna, Via Zamboni, 33, 40126 Bologna, Italy}
\affil[e]{INAF - Osservatorio Astronomico di Brera, via Brera 28, 20121 Milano, Italy}
\affil[f]{ADaptive Optics National laboratory in Italy (ADONI)}

\cftpagenumbersoff{figure}
\cftpagenumbersoff{table} 
\begin{document} 
\maketitle

\begin{abstract}
The ESO Extremely Large Telescope (ELT) will offer unprecedented sensitivity and resolution in the near-infrared, marking a new era for ground-based astronomy. Among its key imaging instruments is MORFEO coupled with MICADO.

MORFEO (Multi-conjugate adaptive Optics Relay For ELT Observations), formerly known as MAORY, is the largest astronomical adaptive optics system ever designed. It features 12 wavefront sensors and three deformable mirrors, for a total of over 20,000 subapertures and over 6,000 actuators. MORFEO represents one of the greatest upcoming challenges in the field of astronomical adaptive optics.

While the design builds upon the heritage of previous AO systems, several architectural choices are entirely new, driven by the unique scale and requirements of this instrument. One of the main challenges is delivering high and uniform wavefront correction across the MICADO field of view. To meet this goal, the MORFEO control strategy adopts a specific approach: sodium laser guide stars are used to sense modes above focus only, since differences in beacon altitude can introduce significant aberrations. Natural guide stars are instead employed to measure and correct for tip, tilt, plate scale variations, and field-averaged focus.

In this work, we present the MORFEO control strategy and provide performance estimates across different observing scenarios.
\end{abstract}

% Include a list of up to six keywords after the abstract

% Include email contact information for corresponding author
{\noindent \footnotesize\textbf{*}Guido Agapito,  \linkable{guido.agapito@inaf.it} }

%\begin{spacing}{2}   % use double spacing for rest of manuscript
\begin{spacing}{1}   % use single spacing 

\section{INTRODUCTION}
\label{sec:intro} 

The ESO Extremely Large Telescope (ELT\cite{2024Msngr.192....3C}) will be the first optical telescope of the next generation to be completed.
It will offer unprecedented sensitivity and resolution in the near-infrared, marking a new era for ground-based astronomy.
Several instruments are in development for this telescope: multi-adaptive optics camera for deep observations (MICADO)\cite{2021Msngr.182...17D}, METIS\cite{2021Msngr.182...22B}, MORFEO\cite{2021Msngr.182...13C}, HARMONI\cite{2021Msngr.182....7T}, ANDES\cite{2021Msngr.182...27M}, and MOSAIC\cite{2021Msngr.182...33H}.

MORFEO (Multi-conjugate adaptive Optics Relay For ELT Observations), formerly known as MAORY\cite{2024SPIE13097E..22C}, is the largest astronomical adaptive optics (AO) system ever designed\cite{BusoniAO4ELT8}.
Its main objective is to provide high resolution images to MICADO\cite{2024SPIE13096E..11S} and HARMONI\cite{2024SPIE13096E..14T} on most of the sky, from the dense region across the galactic plane to the remote regions of the south galactic pole and the HST deep fields\cite{1996AJ....112.1335W,2006AJ....132.1729B}.
It features 12 wavefront sensors and 3 deformable mirrors, for a total of over 20,000 subapertures and over 6,000 actuators.
MORFEO represents one of the greatest upcoming challenges in the field of astronomical adaptive optics and based on the current schedule MORFEO will be the first MCAO system for an ELT that will go on sky.

While the design builds upon the heritage of previous AO systems, several architectural choices are entirely new and are driven by the unique scale and requirements of this instrument.
One of the main challenges lies in delivering high and uniform wavefront correction across the MICADO field of view.
In median atmospheric conditions, this requires a K-band Strehl ratio (SR) of at least 44\% over a field-of-view of 60 arcsec (arcseconds) diameter with a K-band FWHM (Full-Width-Half-Maximum) of no more than 12.7 mas (milliarcseconds) and K-band ensquared energy of at least 30.1 \% in 16 $\times$ 16 mas.
In order to meet the stringent performance requirements of the ELT, MORFEO uses a sophisticated control strategy combining high-order tomographic reconstruction with specialized low-order and phasing loops.

In terms of control strategy, MORFEO shares structural similarities with MAVIS (MCAO-Assisted Visible Imager and Spectrograph\cite{2021Msngr.185....7R}), the Multi-Conjugate Adaptive Optics system designed for the ESO 8-meter Very Large Telescope (VLT).
Both systems rely on tomographic reconstruction using a constellation of Laser Guide Stars (LGSs) for high-order correction.
However, MORFEO is a considerably more complex system due to the larger scale of the ELT, and, crucially, it includes three dedicated Reference WFSs (\textcolor{black}{Ref} WFSs) to provide precise low-order \textcolor{black}{truth sensing (i.e., acting as a slow, un-biased reference)} and Non-Common Path Aberration (NCPA) bias calibration.
%MAVIS, operating on the 8-m VLT, does not include these \textcolor{black}{Ref} WFSs in its baseline design, differentiating the final architecture and control chain.
\textcolor{black}{Unlike MORFEO, MAVIS does not include a dedicated Reference WFS. This is primarily because MAVIS operates on an 8m-class telescope (where LGS spot truncation effects are significantly less severe) and benefits from an integrated imager that simplifies NCPA calibration. In MORFEO, the Ref WFS is the chosen solution to guarantee high stability of the NCPA correction between the LGS sensing path and MORFEO output focal plane, while the internal NCPA of the downstream instruments will be measured by the instruments themselves.}

\textcolor{black}{This work focuses on the system design optimized for MICADO. Specifically, the NGS WFSs \cite{2024SPIE13097E..50B} discussed here are tailored for this instrument. In contrast, while HARMONI will also utilize the MORFEO relay and LGS WFSs, it will integrate a different set of NGS WFSs tailored to its own specific requirements \cite{2024SPIE13100E..7DE}.}

Please note that this work does not describe the system in full detail.
For this, please refer to Busoni et al. 2026~\cite{BusoniAO4ELT8}.
\textcolor{black}{However, to aid readability and provide immediate context, a summary of the key opto-mechanical parameters of the system is provided in Table~\ref{tab:system_parameters}.}
This work also does not report the system estimated performance; this topic is covered in \textcolor{black}{Agapito et al 2026.~\cite{AgapitoAO4ELT8}}.
The MORFEO control system is too complex to be covered in full in a single paper, so this work does not provide all the details.

Several auxiliary loops\textcolor{black}{------i.e., secondary control loops dedicated to managing slow opto-mechanical drifts, maintaining system alignment, and handling spatial or temporal offloads (such as LGS jitter compensation, pupil stabilization, and sodium profile tracking)-----} are described in \textcolor{black}{Busoni et al. 2023~\cite{2023aoel.confE.129B}}.

\begin{table}[ht]
    %\color{blue}
    \centering
    \caption{Summary of key MORFEO system parameters (adapted from Busoni et al. 2026\cite{BusoniAO4ELT8}).}
    \vspace{0.2cm}
    \begin{tabular}{ll}
        \hline\hline
        \textbf{Parameter} & \textbf{Value} \\
        \hline
        \multicolumn{2}{c}{\textit{General System}} \\
        Supported Instruments & MICADO, HARMONI \\
        Science Field of View & Up to $53 \times 53$ arcsec \\
        Technical Field of View (Guide Star patrol) & 160 arcsec diameter \\
        \hline
        \multicolumn{2}{c}{\textit{Deformable Mirrors}} \\
        M4 (Telescope Ground Layer) & $\sim$4500 usable actuators (5352 total) \\
        DM1 ($\sim$17.5 km conjugation) & 1026 actuators \\
        DM2 ($\sim$6.5 km conjugation) & 1147 actuators \\
        \hline
        \multicolumn{2}{c}{\textit{Laser Guide Star Wavefront Sensors (6 units)}} \\
        Sub-apertures & $68 \times 68$ \\
        Field of View & 16.1 arcsec \\
        Pixels per sub-aperture & $14 \times 14$ \\
        Operating Frame Rate & 500 Hz \\
        \hline
        \multicolumn{2}{c}{\textit{Natural Guide Star Low-Order \& Reference Wavefront Sensors (Up to 3 units)}} \\
        \textbf{Reference Arm} & \\
        Sub-apertures & $8 \times 8$ \\
        Field of View (after field stop) & 4.6 arcsec \\
        Pixel Scale & 165 mas/pix \\
        Operating Frame Rate & 10 -- 100 Hz \\
        \rule{0pt}{3ex}\textbf{Low-Order Arm} & \\
        Sub-apertures & 1 (Full-Aperture) or $2 \times 2$ (Shack-Hartmann) \\
        Field of View (after field stop) & 1.8 arcsec (Full-Aperture) / 1.8 arcsec ($2 \times 2$) \\
        Pixel Scale & 7 mas/pix (Full-Aperture) / 15 mas/pix ($2 \times 2$) \\
        Operating Frame Rate & 100 -- 1000 Hz \\
        \hline\hline
    \end{tabular}
    \label{tab:system_parameters}
\end{table}

The paper is organized as follows.
Sec.~\ref{sec:overview} presents an overview of the MORFEO control system, detailing the function and assignment of the various Wavefront Sensor (WFS) types.
\textcolor{black}{Sec.~\ref{sec:slopec} analyzes the critical process of Natural Guide Star (NGS) acquisition, particularly the advanced technique used to achieve the necessary accuracy for faint stars, and subsequently details the Wavefront Sensor Slope Computation methods, highlighting the difference between the dynamic treatment for LGSs and the fixed-window approach for NGSs.}
The following section, Sec.~\ref{sec:focus} focuses on the Average Field Focus Control Strategy and the rationale for managing the focus signal, especially from the LGSs, while Sec.~\ref{sec:truth} presents the truth sensing strategy to compensate the LGS bias due to NCPA and truncation effects.
The temporal control optimization is described in Sec.~\ref{sec:int_time} comprising Sec.~\ref{sec:tt_filter} where the design of the highly effective tip-tilt Temporal Filter (IIR filter) is detailed together with ongoing optimization approaches for gain and control law.
The following section, Sec.~\ref{sec:comm_mat} describes the reconstruction approach, including the Command Matrix and the allocation of commands to the Deformable Mirrors (DMs), Sec.~\ref{sec:pup_control}, focus on WFS pupil image control and Sec.~\ref{sec:update} deals with the update rate and the reference coordinate system.
Finally, the conclusions are drawn in Sec.~\ref{sec:conclusion}.
Furthermore, the appendices provide detailed mathematical frameworks supporting the control strategy: Appendix~\ref{sec:iir_tt_app} details the tip-tilt IIR filter design, Appendix~\ref{sec:multirate_lti} formalizes the multirate sensor fusion, and Appendix~\ref{sec:polc_ss} analyzes the steady-state rejection within a Pseudo Open Loop Control approach.

%%%%%%%%%%%%%%%%%%%%%%%%%%%%%%%%%%%%%%%%%%%%%%%%%%%%%%%%%%%%%%%%%%%%%%%%%

\section{MORFEO Control System Overview}
\label{sec:overview}

\begin{figure}[ht]
    \centering
    \includegraphics[width=0.9\linewidth]{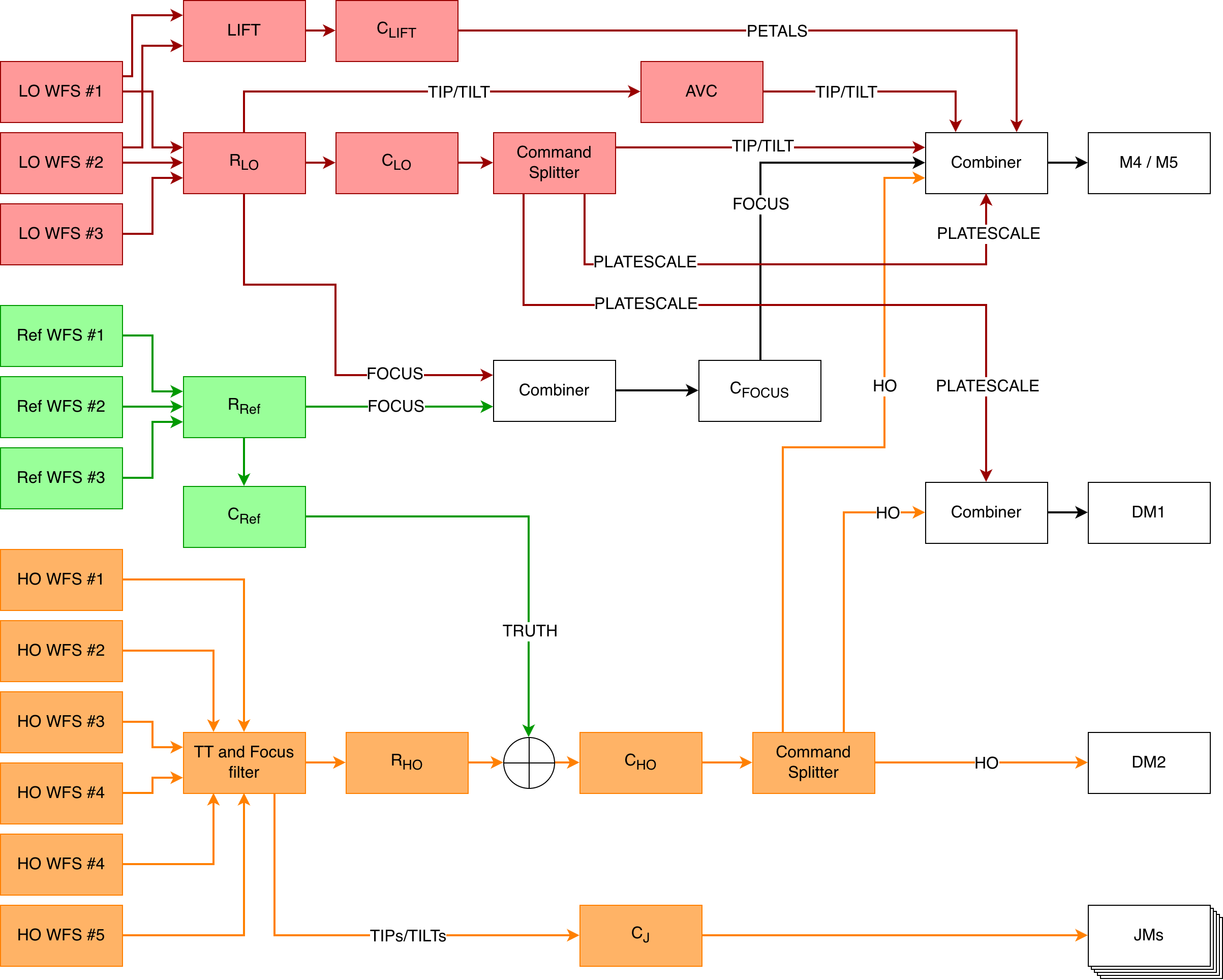}
    \caption{The MORFEO control system diagram. Legend: LO -- Low Orders (IR sensors), Ref -- Reference (truth sensing, VIS sensors), HO -- High Orders (LGS), R - reconstruction/command matrix, C -- temporal controller, TT -- tip-tilt, DM -- deformable mirror, JM -- jitter mirror, LIFT -- Linearized Focal-plane Technique (petalometer), AVC -- Adaptive Vibration Cancellation. Notes: a single AVC block is present in the LO part, but multiple AVC instances can be activated on LO and HO parts.}
    \label{fig:MORFEO_control}
\end{figure}

The \textbf{MORFEO} control system, depicted in Fig.~\ref{fig:MORFEO_control}, employs three distinct sets of Wavefront Sensors (WFSs) to perform the atmospheric turbulence compensation on the science Field-of-View (FoV):
\begin{itemize}
    \item \textbf{Laser Guide Star (LGS) WFSs:} These comprise six Shack-Hartmann Sensors (SHS), each with $68 \times 68$ sub-apertures (more details on the number of sub-aperture choice can be found in Fusco et al. 2022\cite{2022JATIS...8b1514F}). They operate using the sodium return light (589 nm) and are the primary high-order turbulence sensors.
    \item \textbf{Low Order (LO) WFSs:} This set includes two full-aperture WFSs and one $2 \times 2$ SHS. They utilize light from Natural Guide Stars (NGSs) in the J and H infrared bands and are primarily dedicated to sensing global low-order modes \textcolor{black}{(namely overall tip-tilt and focus, with the latter being sensed by the $2 \times 2$ SHS). These measurements are subsequently used to reconstruct both pupil-plane aberrations and field-dependent modes like tilt anisoplanatism, as detailed in the split tomography description below.}    
    \item \textbf{Reference (\textcolor{black}{Ref}) WFSs:} These are three SHS, each with $8 \times 8$ sub-apertures. They work with NGS light in \textcolor{black}{the R- and I-bands} and mainly serve as truth sensors.
\end{itemize}
Please note that the computation of the slope measurements is described in Sec.~\ref{sec:slopec} and that the multi-rate control, which is used on the LO and \textcolor{black}{Ref} to cope with the actual NGS magnitude, is described in Sec.~\ref{sec:multi_rate}.
The WaveFront (WF) reconstruction is performed \textbf{independently} for each of these three sensor sets. A \textbf{tomographic approach} is central to the LGS and \textcolor{black}{Ref} WFSs, enabling the retrieval of the stratification of atmospheric turbulence across a large Field of View (FoV).\\
The reconstruction is performed over a finite number of atmospheric layers — ten, to be precise.
For each layer, a specific number of modes are reconstructed, with this number decreasing with altitude (from approximately 5000 to 2000).
This choice is driven by the fact that:
\begin{itemize}
    \item \textcolor{black}{The tomographic error is minimized near the ground (pupil plane). Moreover, as described by the generalized fitting error framework\cite{2000SPIE.4007.1022R}, the depth of field of an MCAO correction is inversely proportional to the spatial frequency.}
    \item \textcolor{black}{The DM with the highest actuator count is M4\cite{2019Msngr.178....3V}, which is conjugated to about 600m, while the Post-Focal DMs conjugated to high altitudes operate at a lower spatial resolution. Consequently, reconstructing a very large number of modes for high-altitude layers cannot be effectively projected at altitude and would suffer from severe generalized fitting error if projected from near the ground, making it both physically ineffective and computationally inefficient.}
\end{itemize}
Each WFS set is strategically assigned to measure and reconstruct specific modes, following the so-called split tomography approach\cite{2008JOSAA..25.2427G}:
\begin{itemize}
    \item \textbf{LGS WFSs:} These sensors measure all modes above focus (tip-tilt and focus are filtered from the slopes). They are used to reconstruct modes starting from astigmatism in the ground layer, and from the third radial order (Zernike) for the higher-altitude layers. \textcolor{black}{As confirmed by our end-to-end simulations and standard linear systems analysis (e.g., Negro 1984~\cite{1984ApOpt..23.1921N}), filtering these lowest-order modes does not degrade the tomographic reconstruction. The system remains non-degenerate: for instance, the 6 LGSs provide 12 independent measurements of astigmatism, which are mathematically sufficient to uniquely estimate both the pupil-plane astigmatisms and higher-altitude modes.} The filtered Tip-tilt is used to drive the jitter mirrors that correct for the laser position on sky\textcolor{black}{, while the filtered focus is used to drive the LGS WFS focus stages to track the mean altitude of the sodium layer.}   
    \item \textbf{LO WFSs:} These measure tip-tilt and, via the $2 \times 2$ SHS, also Focus. These measurements are used to reconstruct:
    \begin{itemize}
        \item Tip-Tilt on the ground/pupil.
        \item \textcolor{black}{Plate scale} modes (linear distance variations between stars in the field).
        \item Field rotation.
        \item The average Focus across the field.
    \end{itemize}
    \item \textbf{\textcolor{black}{Ref} WFSs:} These sensors measure approximately 50 low-order modes, starting from tip-tilt. They are used to reconstruct these $\approx 50$ modes across all atmospheric altitudes.
\end{itemize}
There is a deliberate \textbf{overlap} between the three WFS sets, as the \textbf{\textcolor{black}{Ref} WFSs} function as \textbf{truth sensors}\textcolor{black}{---slow wavefront sensors operating on Natural Guide Stars that measure and compensate for low-frequency drifts and LGS-specific artifacts (such as sodium layer variations) that the fast LGS WFSs cannot disentangle from turbulence.}
Their measurements are used to remove low-order bias from the LGS WFS measurements, correcting for effects such as Non-Common Path Aberrations (NCPA) and WFS truncation errors as described in Sec.~\ref{sec:truth}.
Furthermore, the average focus across the field is measured by both the LO and \textcolor{black}{Ref} WFSs as described in Sec.~\ref{sec:focus}.
The combination of these two focus measurements can be used to reduce anisoplanatism effects (the degradation of correction quality away from the guide star).

The three primary Wavefront Sensor (WFS) sets (LGS, LO, and \textcolor{black}{Ref}) are inherently insensitive to phasing errors that may arise from the segmentation of the telescope primary mirror and the segmented M4 Deformable Mirror (DM).
These errors must be addressed because they lead to common aberrations known as petal modes.
To detect and correct these specific segmentation-induced errors, the MORFEO system incorporates dedicated \textbf{focal plane sensors}.
These sensors are integrated into the full-aperture LO WFSs by generating a second out-of-focus spot using a narrow-band notch filter.
The algorithm selected for the reconstruction of these phasing errors is the \textbf{LIFT (Linearized Focal-plane Technique)} \cite{2010OptL...35.3036M}.
LIFT is specifically employed to measure and correct the petal modes associated with the segmented structure of the telescope and M4 and it is described in Plantet et al. 2026\cite{PlantetAO4ELT8}.

A significant departure from the initial design exploration of the petalometer\cite{2022SPIE12185E..56A} is the transition from focal plane sensing within the sub-apertures of a $2 \times 2$ SHS to the current approach, which uses a dedicated out-of-focus spot.
This evolution brings several key advantages:
\begin{itemize}
    \item The dedicated out-of-focus spot approach provides significantly improved sensitivity to the phasing errors.
    \item Essentially, this design solves a major limitation of the previous $2 \times 2$ SHS concept: the inherent blindness that occurred when the telescope spider arms were aligned with the sub-aperture discontinuities of the sensor. The new dedicated sensor avoids this geometrical constraint entirely.
    \item The dedicated out-of-focus spot is produced with a narrow band filter so greatly reducing the chromatism and the problem related to it, like the polychromatic smearing of PSFs (see Kuznetsov et al. 2024 \cite{2024A&A...687A.221K}).
\end{itemize}

After reconstruction, the LGS and \textcolor{black}{Ref} WFS signals are combined to remove the low-order bias. \textbf{Temporal filters} are then applied to generate the final commands for the Deformable Mirrors (DMs).
An integrator controller is the standard choice, except for tip-tilt, where a more complex filter is implemented.
Tip-tilt are critically affected by large disturbances due to \textbf{wind shake} on the ELT M1 and M2 mirrors and \textbf{vibrations}. The specific tip-tilt filter is detailed in Sec. \ref{sec:tt_filter}. \\
In parallel to the temporal filters there is the \textbf{Adaptive Vibration Cancellation (AVC)} \cite{7320616} algorithm: This algorithm is used to actively reject narrow-band vibration peaks. It can be enabled for multiple modes and frequencies as required by the system.

Finally, the reconstructed modes are distributed into three sets, one for each DM. Specifically, the \textbf{Focus and Astigmatism} commands for M4 are a combination of the ``standard'' commands and the \textbf{\textcolor{black}{plate scale} commands}. This is necessary because ground-layer Focus and Astigmatisms must be precisely counter-compensated on the post-focal DMs to produce a pure \textcolor{black}{plate scale} mode without altering the overall average Focus and Astigmatisms in the field.
\textcolor{black}{Conversely, for the post-focal DMs, the combiner does not mix the LO and LGS commands for these low-order modes, but rather maps them onto strictly distinct, separate modes.}

Please note that all of this \textcolor{black}{will be} implemented in MORFEO's instrument control software and real-time control, as described in \textcolor{black}{Costa et al. 2026~\cite{CostaAO4ELT8} and Capasso et al. 2024~\cite{2024SPIE13101E..3XC}}.

%%%%%%%%%%%%%%%%%%%%%%%%%%%%%%%%%%%%%%%%%%%%%%%%%%%%%%%%%%%%%%%%%%%%%%%%%

\section{\textcolor{black}{Natural Guide Star Acquisition and} Wavefront Sensor Slope Computation}\label{sec:slopec}

The accurate calculation of the wavefront slopes from the WFS focal plane images is critical and differs slightly between the LGS and NGS sensors due to their distinct signal characteristics. The primary method used for slope calculation is the Center of Gravity (CoG) algorithm, implemented with specific windowing and weighting strategies for optimal performance\cite{2006MNRAS.371..323T}.

\subsection{Natural Guide Star Acquisition}
\label{sec:ngs_acq}

The success of the MORFEO Multi-Conjugate Adaptive Optics (MCAO) system hinges on the accurate and prompt acquisition of Natural Guide Stars (NGSs).
NGSs are essential for measuring and correcting low-order modes such as tip-tilt, average field focus, and platescale variations, as the LGS WFSs are filtered to exclude these modes as described in Sec.~\ref{sec:overview}.

The acquisition sequence (already detailed in Agapito et al. 2023\cite{2023aoel.confE..10A}) ensures that the NGS spots are correctly centered on their respective Wavefront Sensors (WFSs) for closed-loop operation:

\begin{itemize}
    \item Based on the scientific target, up to three suitable NGSs are selected to feed the three LO and \textcolor{black}{Ref} WFSs.
    \item The telescope and the MORFEO internal stages are commanded to place the selected NGSs into the fields of view of the \textcolor{black}{Ref} WFSs while the telescope is pointing.
    \item \textcolor{black}{The NGS acquisition can be performed in a seeing-limited regime (relying exclusively on the telescope active optics to mitigate large telescope aberrations) or after the High Order (HO) loop has been closed (where the system is partially corrected). The former is currently being evaluated as the preferred baseline, as it avoids potential low-order biases introduced by the LGS HO loop and enables a direct, single-step transition from telescope control to the combined HO and Low Order (LO) control regime.}
    \item The \textcolor{black}{Ref} WFS is used for the first step because it offers a larger FoV ($4.6 \, \text{arcsec}$) compared to the target $\text{FoV}$ of the LO WFS ($1.8 \, \text{arcsec}$).
    \item The accuracy of this initial step must be sufficient to move the star into the smaller LO WFS $\text{FoV}$. \textcolor{black}{Because the exact location of the star across the detector is initially unknown, this dedicated acquisition phase on the Ref WFS requires an integration time of $1\,\text{s}$. This value naturally differs from the much faster closed-loop tracking parameters listed in Tab.~\ref{tab:r_int_time}.} This speed presents a challenge, especially with dim stars (limiting magnitude $R=21.5$) in the presence of moonlight.
\end{itemize}

\begin{figure}[h]
    \centering
    \includegraphics[width=0.8\linewidth]{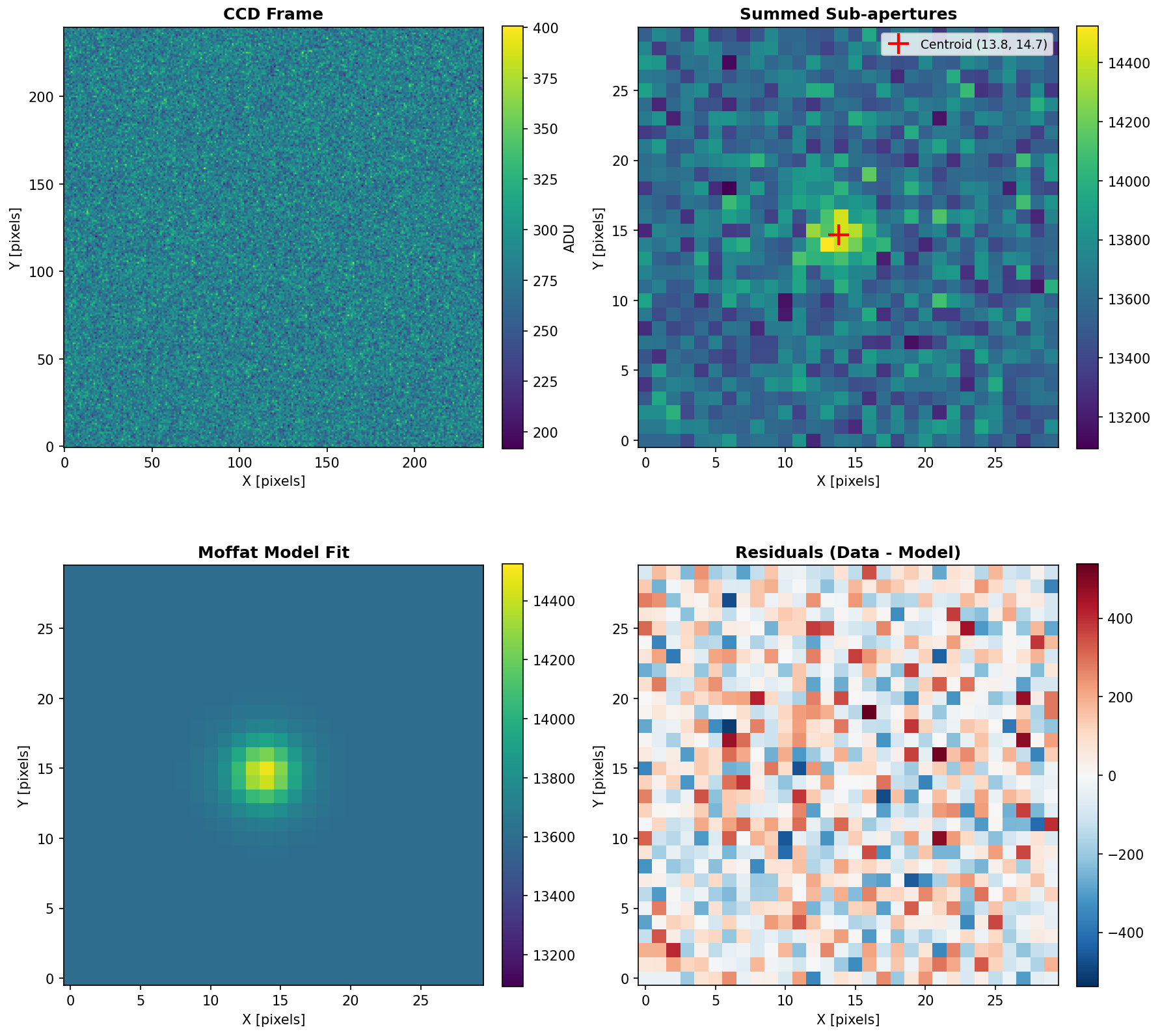}
    \caption{Simulation results for an R=21.5 star during the acquisition phase (1s integration time) with sky background from the Moon. Top-left: full CCD frame where the spot is hidden by noise. Top-right: sum of all valid sub-apertures, revealing the NGS spot. Bottom-left: Moffat model fitting. Bottom-right: fitting residuals.}
    \label{fig:fitting_results}
\end{figure}

An initial analysis of the acquisition process (as detailed in Agapito et al. 2023\cite{2023aoel.confE..10A}) proved to be overly pessimistic.
This pessimism was overcome by anticipating an advanced technique, similar to that employed by ESO at the VLT\cite{berdeu_2026_21104330}, which is now implemented in the MORFEO design.

\begin{itemize}
    \item This method involves combining frames or portions of frames from all valid sub-apertures to significantly enhance the signal-to-noise ratio.
    \item \textcolor{black}{This technique is effective because, in the visible band, the spot size is dominated by the seeing halo (typically larger than 0.5 arcsec). Differential low-order aberrations between sub-apertures are significantly smaller than this angular scale. Thus, when the frames are co-added, the spots strongly overlap, producing a detectable "mega-spot." While not diffraction-limited, this combined spot provides sufficient signal-to-noise ratio to guarantee the coarse centroiding accuracy required to center the star.}
    \item Using this technique on an $R=21.5$ star under $1.1 \, \text{arcsec}$ line-of-sight seeing and high sky background flux (e.g., from the Moon at $30^\circ$ from the target), the estimated $\text{RMS}$ error is reduced to $0.5 \, \text{pixel}$ with $1 \, \text{s}$ of integration. This corresponds to an angular error of $80 \, \text{mas}$ (see Fig.~\ref{fig:fitting_results}), which is much better than achievable with a classical tip-tilt estimation and is comfortably below the required accuracy for the LO WFS $\text{FoV}$ of $1.8 \, \text{arcsec}$.
    The numerical evaluation of these acquisition procedures, including the 2D Moffat fitting, flux estimation, and centroid tracking on noisy focal plane images, is performed within the \texttt{SPECULA} environment using \textcolor{black}{a dedicated spot monitoring routine}.
\end{itemize}

Accurate NGS acquisition is therefore the indispensable first step for transitioning from open-loop observation to stable, high-performance MCAO closed-loop operation.

\subsection{LGS WFS Slope Calculation}

For the six LGS WFSs, the high variability in the sodium layer flux necessitates a dynamic approach to manage the spot signals\cite{obertiAO4ELT6}:

\begin{itemize}
    \item The slopes are computed using the CoG algorithm.
    \item A specific window (or region of interest) is employed, with its size and position determined by the average intensity of the sodium return light.
    \item The window size is dynamically set to prevent truncation of the LGS spot during periods of high flux, while simultaneously minimizing the noise contribution from unilluminated pixels. This balances the need to capture all signal energy against noise propagation.
\end{itemize}

A relevant additional point is that in cases where laser propagation is fully inhibited (e.g., due to aircraft passage or interlocks), the RTC injects zero slopes into the control loop. This strategy prevents the high-order integrator from diverging, effectively holding the deformable mirror shape or allowing a controlled relaxation, depending on the specific loop configuration.

\subsection{NGS WFS Slope Calculation (LO and \textcolor{black}{Ref})}
\label{sec:ngs_slope_calc}

The Natural Guide Star (NGS) sensors (LO and \textcolor{black}{Ref} WFSs) exhibit a spatially uniform spot morphology compared to their LGS counterparts, allowing for a standardized, rigid-window approach. However, the LO WFS presents a specific geometrical challenge: it features a very fine pixel scale over a large Field of View (encompassing approximately 200 pixels across in the full-aperture configuration), while the optimal computation window or weighting map is restricted to only a few pixels to minimize detector noise propagation (one pixel is approximately $\lambda / D$, where $\lambda$ is the central sensing wavelength, $\approx$ 1400 nm).

To accurately place this small region of interest, especially during the initial loop closure transients, the system employs a two-step Center of Gravity (CoG) strategy:

\begin{itemize}
    \item To locate the spot within the large uncorrected FoV, the system relies on \textcolor{black}{a continuous spot monitoring process} (see Sec.~\ref{sec:ngs_acq}). By averaging a short rolling buffer of frames (typically around ten, dynamically adjusted based on the instantaneous Signal-to-Noise Ratio), the module performs a robust 2D profile fitting to determine the precise global centroid coordinates. Crucially, during this initial transient phase, the computation window is intentionally initialized with a larger footprint to reliably capture the partially corrected spot. As the loop converges and the spot sharpens, the window is progressively shrunk to its final, optimal dimensions over a short period (a few tenths of seconds).
    \item Once the closed-loop stabilizes and the spot is firmly confined within the small tracking window, the window's position is frozen (or updated extremely slowly). Within this static region, the standard weighted CoG is computed at the maximum frame rate, preventing statistical noise from inducing artificial window jitter.
    \item Once steady-state is reached, \textcolor{black}{the size and shape of the tracking window are strictly fixed. This rigid steady-state approach differs from the dynamic LGS processing (which continuously adapts its windows based on average sodium layer variations) and is permissible due to the highly stable morphological behavior of NGS spots compared to LGS spots, which suffer from sodium layer altitude and thickness variations.} However, to ensure system robustness, \textcolor{black}{time-averaged frames are continuously monitored}. If the spot is temporarily lost or severely drifts due to unexpected anomalies (e.g., passing clouds or extreme windshake), this background monitoring detects the failure and can immediately trigger a re-centering or re-acquisition sequence.
\end{itemize}

%%%%%%%%%%%%%%%%%%%%%%%%%%%%%%%%%%%%%%%%%%%%%%%%%%%%%%%%%%%%%%%%%%%%%%%%%

\section{Average Field Focus Control Strategy}
\label{sec:focus}

The control of the average focus across the field presents significant challenges for Laser Guide Stars base system for Extremely Large Telescopes (ELTs) due to the large aperture.
For smaller, $8 \, \text{m}$-class telescopes, LGS Wavefront Sensor (WFS) measurements are typically used for focus sensing, with the focus signal unbiased using a Natural Guide Star (NGS) truth sensor to compensate for the unknown and variable altitude of the sodium layer.

\begin{figure}[h]
    \centering
    \includegraphics[width=0.5\linewidth]{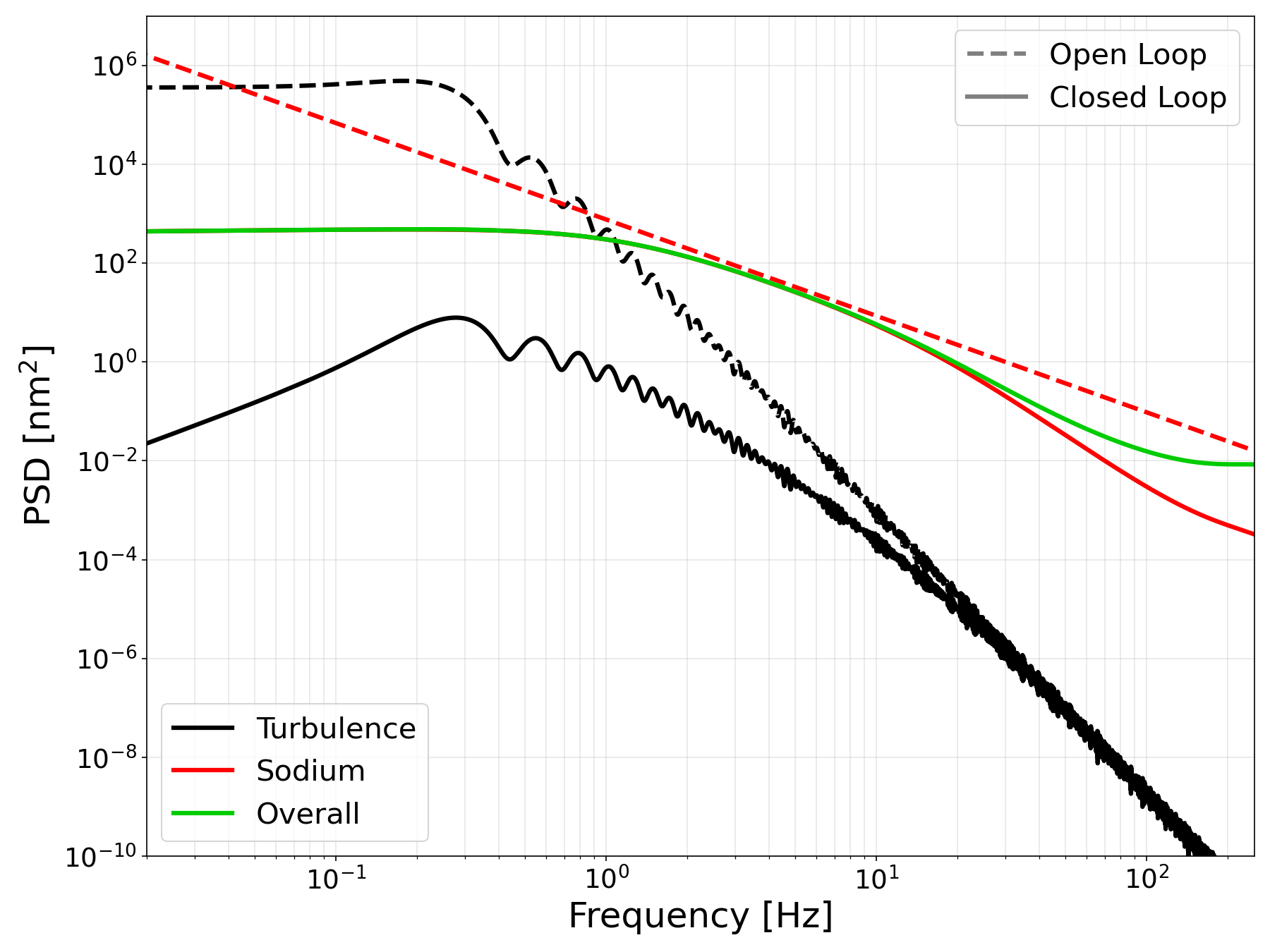}
    \caption{
    \textcolor{black}{Power Spectral Densities (PSDs) of the focus term. Dashed lines represent the open-loop PSDs, while solid lines denote the closed-loop residual PSDs. The closed-loop transfer functions assume an integral controller with a gain of 0.75 for the NGS and 0.25 for the LGS. The plot illustrates how the focus disturbance is dominated by sodium variations at low frequencies and atmospheric turbulence at higher frequencies.}
    }
    \label{fig:focusPSDs}
\end{figure}

However, the transition to an ELT with a diameter ($D$) approximately five times larger dramatically impacts the conversion factor between the sodium layer's altitude variations and the induced focus aberration. The relationship is governed by the equivalent focal length of the telescope, leading to a scaling factor proportional to the square of the aperture ratio, i.e., $(D_{\text{ELT}} / D_{8\text{m}})^2 \approx 25$.
This factor means that the temporal dynamics of the sodium layer altitude variations induce a focus aberration that is significantly \textbf{stronger} than the aberration caused by atmospheric turbulence itself. This dominance is clearly illustrated in the Power Spectral Densities (PSDs) shown in Figure \ref{fig:focusPSDs}, where the residual focus error is observed to be largely dictated by the sodium layer induced component.

\textcolor{black}{From a control theory perspective, it is essential to distinguish how these two phenomena propagate. Atmospheric turbulence acts as an external physical disturbance that the AO loop mitigates via its disturbance rejection transfer function. Conversely, sodium altitude variation acts as a sensor measurement bias. If the LGS focus were left in the main control loop, the system would mistakenly track this false signal, injecting it directly onto the deformable mirrors via the closed-loop tracking transfer function. Because the closed-loop response is approximately unity at low frequencies, this large sodium-induced error would be fully printed onto the science wavefront. This fundamental difference dictates that the LGS focus signal must be actively filtered out.}

Another critical factor is the \textbf{differential focus} between the multiple LGS WFSs, which arises from horizontal non-uniformities in the sodium profile across the field\cite{2013MNRAS.432L..21R}.
The system exhibits an extreme sensitivity to this effect: \textcolor{black}{we estimate that a differential sodium altitude of only $10 \, \text{m}$ across the six LGS beams at a zenith angle of $30^\circ$ results in a wavefront error of approximately $20 \, \text{nm}$ RMS. High-resolution UBC Lidar campaigns have demonstrated that the mesospheric sodium exhibits severe structural differences and strong horizontal anisotropy on spatial scales relevant to ELT asterisms~\cite{2012SPIE.8447E..19P}. Consistent with these complex dynamics, upper limits derived from MCAO on-sky telemetry indicate that transversal altitude variations can reach up to roughly 100 m~\cite{2013MNRAS.432L..21R}. Because actual differential altitudes across the beams can easily span several tens of meters, this effect rapidly scales to large overall wavefront errors reaching hundreds of nanometers. Hence, an uncorrected sodium layer altitude variation translates into an unacceptable degradation of performance.}

\textcolor{black}{It is worth noting that due to the off-axis launch geometry of the LGSs and the finite Field of View of the sub-apertures, variations in the sodium profile do not inject a pure focus term, but also introduce smaller low-order aberrations (e.g., astigmatism). However, while the dominant focus term requires explicit fast detrending, these higher-order artifacts are small in amplitude and are heavily mitigated by the large LGS WFS FoV and the optimized centroiding algorithms\cite{obertiAO4ELT6, 2022JATIS...8b1514F}. Any residual slow variations of these low-order modes are effectively compensated by the \textbf{Ref WFS} truth sensing loop, thus not requiring the same explicit fast filtering.}

\textcolor{black}{Hence, the adopted control strategy dictates that both the average and the differential field focus must be controlled \textbf{using exclusively Natural Guide Star (NGS) measurements}. Therefore, the focus signal derived from the LGS WFSs must be actively \textbf{filtered out} from the main control loop, similar to the approach used for the Tip-Tilt signals.}

\textcolor{black}{MORFEO utilizes four WFSs capable of measuring focus based on NGSs: the single} $2 \times 2$ Low Order (LO) WFS operating in the IR band (J and K) and the three $8 \times 8$ Reference (\textcolor{black}{Ref}) WFSs operating in the visible and near-IR band (R and I).

These sensors have complementary characteristics:
\begin{itemize}
    \item The LO WFSs provide the best signal-to-noise ratio for focus measurements, especially in the presence of strong sky background light, such as moonlight, but they suffer from a reduced linearity range.
    \item \textcolor{black}{The Ref WFSs possess a wider linearity range for global focus measurements. Because their sub-apertures are smaller relative to the pupil diameter, the intra-subaperture phase variation induced by a global focus is significantly reduced. This keeps the focal plane spots compact and preserves the centroiding linearity over a wider dynamic range\cite{1984ApOpt..23.1921N}. In addition, having three Ref WFSs available allows for focus measurement on multiple lines of sight, which can reduce the remaining anisoplanatic effects.}
\end{itemize}

\textcolor{black}{The wider linearity range of the Ref WFSs is particularly critical during the initial acquisition and loop-closure transients. Once in steady-state, as anticipated in Sec.~\ref{sec:overview}, the resulting strategy is to \textbf{combine the focus measurements} from all four sensors (when three NGSs are available). This fusion remains highly advantageous for two reasons: it reduces the remaining anisoplanatic effects, and it provides robustness for diverse asterism geometries. For instance, if the brightest NGS is allocated to a full-aperture LO WFS to optimize tip-tilt correction, the $2 \times 2$ LO WFS may operate on a fainter star, making the supplemental focus data from the Ref WFSs highly valuable. By weighting each measurement differently based on instantaneous flux, noise levels, and the operating point (with weights dynamically dropping to zero for overly noisy sensors), the system guarantees an overall measurement fidelity that always meets or exceeds the performance of the LO WFS alone (for consideration on different sensing and control rates see Sec.~\ref{sec:multi_rate}).}

%%%%%%%%%%%%%%%%%%%%%%%%%%%%%%%%%%%%%%%%%%%%%%%%%%%%%%%%%%%%%%%%%%%%%%%%%

\section{Truth Sensing}
\label{sec:truth}

The Truth Sensing, also referred to as the Reference loop, is a control loop based on Natural Guide Stars (NGSs). Its primary function is to detrend pseudo-static aberrations introduced by the High Order loop onto the Deformable Mirrors (DMs). These aberrations mainly arise from two sources:
\begin{enumerate}
    \item Non-Common Path Aberrations (NCPA) between the LGS WFSs and the scientific focal plane.
    \item LGS sensing artifacts, including spot truncation and profile-induced centroid biases.
\end{enumerate}

\textcolor{black}{Regarding the second point, LGS sensing artifacts manifest in two primary ways. First, despite the large 16.1 arcsec FoV and dynamic windowing, residual spot truncation can still introduce slowly varying low-order aberrations on the order of several tens of nanometers RMS, as evaluated in our extensive simulation campaigns\cite{obertiAO4ELT6,2020SPIE11448E..2SA,2022JATIS...8b1514F,AgapitoAO4ELT8}. Second, the temporal evolution of the sodium profile introduces additional challenges. As the time-varying vertical structure of the sodium layer maps non-linearly into sub-aperture elongation, it creates shifting spot asymmetries, even in the absence of hard truncation, the Center of Gravity algorithm suffers from centroid biases that map into slowly varying low- and medium-order aberrations. Based on our analysis using high-resolution temporal sodium profile data~\cite{2012SPIE.8447E..19P}, the residual wavefront error specifically related to these temporal variations is approximately 15 nm RMS~\cite{AgapitoAO4ELT8}. Both of these slow-drifting artifacts must be actively detrended.}

The reference loop is responsible for controlling low-order modes starting from astigmatism (as tip-tilt and focus are already controlled by the NGS LO loop) to correct for the reconstruction of aberrations in the LGS objective and these truncation and profile-induced signals.

\textcolor{black}{The quasi-static aberrations measured by the LGS WFSs—and consequently the deformations applied to the DMs—are not strictly constant. These slow drifts depend on external conditions such as the sodium layer altitude, its vertical profile, and the telescope elevation angle, which affects the LGS footprints on the MORFEO optics. While a priori knowledge derived from system models and calibrations allows for an approximate estimation of these NCPAs during an observation, a real-time estimation of the residual pseudo-static aberrations using a Reference WFS is required. However, while it is often assumed that rapidly varying atmospheric turbulence averages out over long integration times, recent studies\cite{2011ApOpt..50.5303G,2026RASTI...5af061G} have demonstrated that on ELT-scale apertures, the finite wind-crossing time leaves low-frequency residuals known as Turbulence-Induced Quasi-Static Aberrations (TIQSA). This physical limitation, combined with the limited spatial resolution ($8\times8$ sub-apertures) of the Reference WFSs, prevents them from perfectly filtering out the volumetric turbulence. Therefore, any attempt to run the Reference loop at higher bandwidths would erroneously inject these atmospheric residuals as static commands. Therefore, the Truth Sensing loop must be operated at extremely low bandwidths to provide a time-averaged measurement as close as possible to the true static wavefront. 
Furthermore, the planned calibration strategy relies on these Reference WFS measurements to continuously update the NCPA look-up tables (LUTs) used for the initial presets. While the detailed calibration procedures will be finalized prior to the completion of the instrument Final Design Review phase, the baseline approach relies on long-term statistical averaging over multiple observations. This temporal averaging effectively filters out zero-mean transient sodium fluctuations, isolating the deterministic optical flexures. Moreover, if a persistent, non-zero mean sodium bias exists, it effectively acts as a static sensing offset that is beneficially captured by the LUT. In any case, the LUT strictly provides the initial preset; during the observation, the real-time Reference loop continuously compensates for any residual discrepancy—whether induced by flexure or transient mesospheric dynamics—ensuring optimal scientific performance.}

A critical challenge in MORFEO is that the NGSs used for the Reference loop are located in the technical patrol field, significantly off-axis compared to the scientific \textcolor{black}{instruments} (MICADO \textcolor{black}{and HARMONI}). Due to telescope and relay optical aberrations, an NGS senses a differential aberration with respect to the scientific \textcolor{black}{field} center.
Hence, driving the AO loop to flatten the wavefront in the direction of the NGSs does not ensure a flat wavefront at the MICADO focal plane.
To address this, the signals from the Reference WFS must be offset by an amount that depends on the specific position of the NGS in the patrol field.
This requires accurate knowledge (via modeling and calibration) of the differential NCPA between the specific NGS position and the scientific field. However, an in-depth discussion of the calibration procedures required to measure and maintain these differential NCPAs—which involve complex interactions with the scientific instruments—falls outside the scope of this paper, which focuses strictly on the control strategy execution.

In the baseline configuration, MORFEO employs a tomographic truth sensing approach using three Reference WFSs to discern correction errors at different altitudes.
This has been analyzed in \textcolor{black}{Busoni et al. 2019~\cite{Busoni2019} and Pariani et al. 2024\cite{2024SPIE13097E..4RP}}.
The general control strategy involves two stages:
\begin{enumerate}
    \item For a given observing condition, the modal deformation induced by the modeled/calibrated LGS NCPA is estimated at preset. These values are set as open-loop offsets to be subtracted from the LGS loop measurements. This step places the system close to the optimal working point, accelerating the convergence of the closed loop.
    \item After the main AO loop is closed, the Reference loop is enabled. This loop operates at a very low bandwidth (approx. 0.01 Hz) to avoid compensating for atmospheric turbulence. \textcolor{black}{Because the system is already placed close to the optimal working point by the preset offsets (Step 1), this low bandwidth does not impose a significant acquisition overhead. The loop only needs to clean up small residual errors, allowing the scientific performance to reach its nominal plateau in a few tens of seconds (typically 20--30 s), well within standard ELT acquisition procedures.} At each iteration, the computed correction is integrated into the previous offset.
\end{enumerate}

Similar to the main AO loop, the reconstructor and projection matrices for the Reference loop are updated dynamically based on the elevation angle, derotation angle, and NGS positions.

\textcolor{black}{Corrections derived from the Truth Sensing loop are incorporated as \textbf{modal offsets} directly into the differential command vector of the LGS High Order (HO) loop.}
The control law is described by:
\begin{equation}
    c(k) = f \cdot \Delta c(k) + r(k)
\end{equation}
where $c(k)$ is the final command vector at time step $k$, \textcolor{black}{$\Delta c(k)$ represents the instantaneous command increments (computed by multiplying the current LGS WFS residual slopes by the command matrix defined in Eq.~\ref{eq:comm_mat}), $f$ is the temporal filter applied to these high-order differential commands}, and $r(k)$ is the reference modal offset. The value $r(k)$ is obtained by integrating the differential commands resulting from the truth sensing reconstruction.
\textbf{Advantages of Modal Offsets:} we deliberately choose to apply offsets in the modal domain rather than as slope offsets for two key reasons:
\begin{itemize}
    \item \textcolor{black}{It avoids an additional Matrix-to-Vector (M2V) multiplication. \textbf{The Truth Sensing loop implementation used in MORFEO} naturally computes a modal vector; converting this into reference slopes would require an explicit multiplication by the LGS interaction matrix.}
    \item \textcolor{black}{The modal offset approach avoids propagating cross-talk errors. If applied as slope offsets, the correction must subsequently pass through the LGS tomographic reconstructor. Because the reconstructor is not the perfect inverse of the interaction matrix (due to regularization and optical mis-calibrations), a pure low-order truth signal would incorrectly map into spurious higher-order modes. Applying the correction directly in the modal domain bypasses this reconstructor step entirely, ensuring a $1:1$ transmission of the truth signal that relies solely on the stable mode-to-command matrix.}
\end{itemize}

%%%%%%%%%%%%%%%%%%%%%%%%%%%%%%%%%%%%%%%%%%%%%%%%%%%%%%%%%%%%%%%%%%%%%%%%%

\section{Temporal Control Optimization}
\label{sec:int_time}

This section details the parameters governing the temporal control of the system, specifically the management of control loop gains, the selection of controlled modes, the integration times, the handling of multi-rate sensors, and the strategy for vibration rejection.

\subsection{Gain Scheduling and Modal Control Strategy}
The control strategy distinguishes between static gains and online optimization:

\begin{itemize}
    \item \textcolor{black}{HO Loop: the control gains are kept constant during standard closed-loop operations. Currently, we are considering a value of 0.25 for all HO modes. Despite the high-flux LGS regime, this relatively low gain is strictly constrained by the $\approx$ 3-frame total system latency to guarantee sufficient stability margins. Specifically, it ensures compliance with the ESO robustness requirement for the closed-loop sensitivity function ($\vert{}\vert{}S(j\omega)\vert{}\vert{}_{\infty} < 2$). In addition, because modal turbulence evolves more slowly on a 39-m ELT aperture compared to 8-m class telescopes~\cite{1995JOSAA..12.1559C}, this conservative gain optimally balances temporal tracking error against noise and aliasing propagation without sacrificing overall performance.}
    
    \item \textcolor{black}{Ref} Loop: similar to the HO loop, the Reference gains are constant during operations. Their values are pre-calculated and selected from a Look-Up Table (LUT) based on the specific NGS R-band magnitude and the selected framerate.
    
    \item LO Loop: this is the only loop where gains are optimized online. This dynamic optimization is critical to maximize the rejection of windshake, which is identified as a major contributor to the overall wavefront error. As detailed in Section \ref{sec:tt_filter}, the LO control employs a complex IIR filter rather than a classical integrator, requiring continuous tuning to match the evolving disturbance environment.
\end{itemize}

\textcolor{black}{Please note that a gain scheduling strategy (ramping up gains) can be implemented during the transition from open-loop to closed-loop. This progressive closure mitigates severe transient effects, preventing large instantaneous commands that could otherwise cause actuator saturation, excite structural resonances, or push the wavefront sensors out of their linear dynamic range.}
Finally, regarding the modal basis, the number of controlled modes is generally fixed (except during the initial loop closure transient).
However, in conditions of poor seeing or low \textcolor{black}{Signal-to-Noise Ratio (SNR)}, the system is designed to update the Pseudo-Open Loop (POL) \cite{2003SPIE.5169..206E} reconstruction matrices to increase the leaky factor of the temporal control (more details on the leaky factor can be found in the Appendix~\ref{sec:polc_ss}).
This effectively regularizes the correction, reducing noise and spatial aliasing propagation on high-order modes, \textcolor{black}{as verified by our end-to-end numerical simulations.}

\subsection{Tip-Tilt Temporal Filter Design}
\label{sec:tt_filter}

The tip-tilt modes are affected by a strong disturbance originating from wind shake and mechanical vibrations on the \textcolor{black}{telescope secondary mirror (M2).} This disturbance is significant, featuring more than $10 \, \mu m$ RMS, with most of its energy concentrated at temporal frequencies lower than about $1 \, \text{Hz}$.
\begin{figure}[ht]
    \centering
    \includegraphics[width=0.5\linewidth]{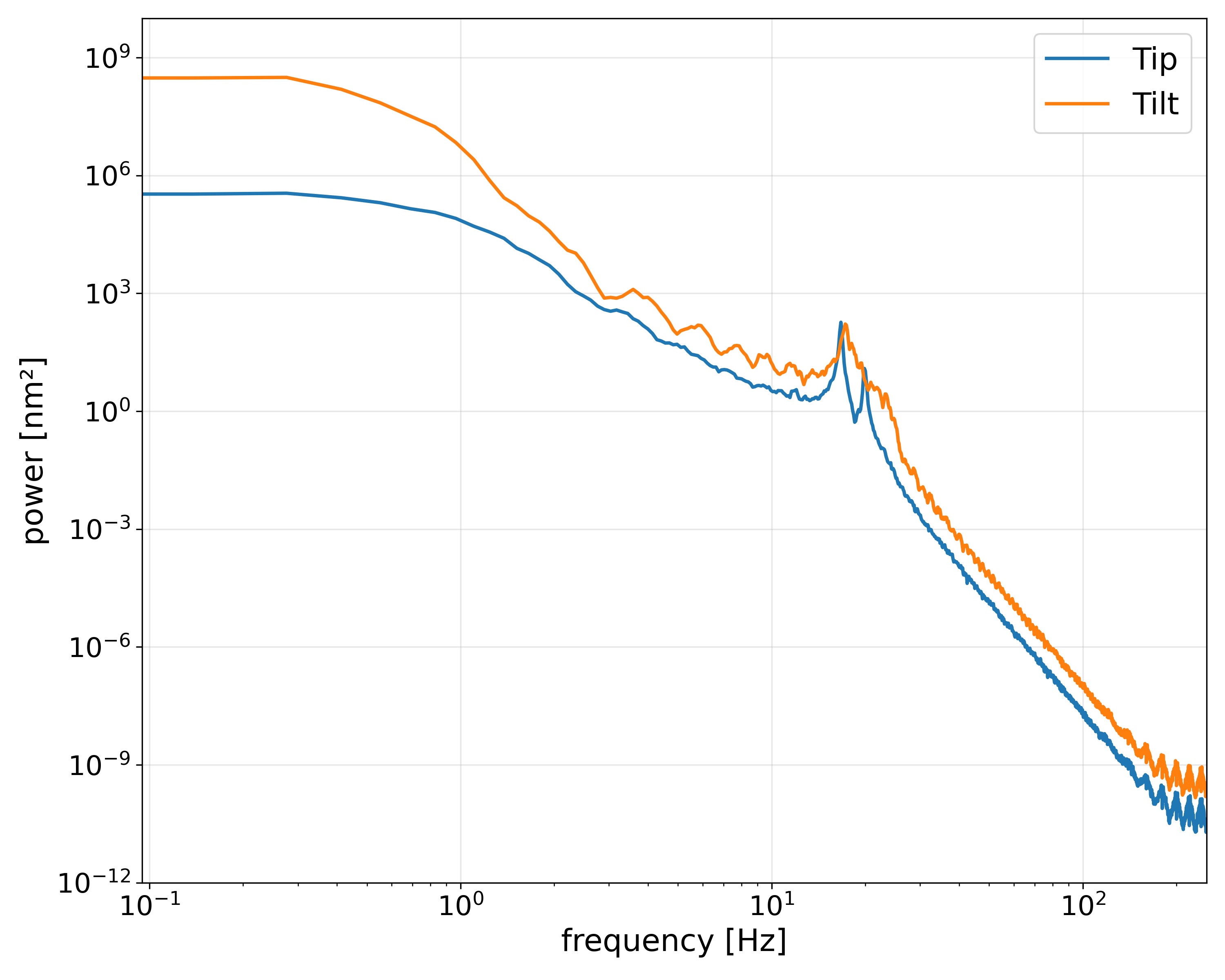}
    \caption{\textcolor{black}{Power Spectral Density (PSD) of the tip-tilt windshake disturbance on the telescope secondary mirror (M2). Data is extracted from the ESO Telescope Wavefront Perturbations Data Package \cite{ESO271292}. The baseline simulation assumes the telescope at a $45^\circ$ zenith angle with the windscreen deployed, subjected to a frontal wind load following a von Kármán pressure model (mean wind speed of 8~m/s at the M2 level).}}
    \label{fig:M2_WS_PSD}
\end{figure}
Following previous foundational works on tip-tilt (TT) control \cite{2012SPIE.8447E..31A,2012OExpr..2027108A,2013aoel.confE..24A}, which established the necessity of high low-frequency disturbance rejection, we designed an \textbf{Infinite Impulse Response (IIR) filter} for the TT control loop, which typically operates at a sampling frequency ($f_s$) of $500 \, \text{Hz}$.
While the previous studies also focused on producing IIR filters to achieve superior disturbance rejection, their specific design approach (e.g., pole-zero placement, filter order, or target rejection bandwidth) differed from the current implementation. Nonetheless, the core strategy remains the same: the use of an IIR architecture, which offers greater flexibility and effectiveness compared to the simple integrator, is essential to effectively mitigate the strong low-frequency disturbance originating from wind shake and M2 vibrations, which dominates the TT error budget.
The IIR structure provides aggressive and superior rejection at low frequencies compared to standard integrators. It is important to note that the final calibration of the filter pole and zero positions will be an empirical process, requiring fine-tuning based on the actual temporal characteristics of the disturbance measured at the telescope site.
The chosen pole and zero approach has been validated as a highly effective baseline design to achieve the required disturbance rejection performance.
More details on the filter can be found in Appendix~\ref{sec:iir_tt_app}.

\subsection{Adaptive Vibration Cancellation}
\label{sec:avc_detail}

As introduced in the system overview (Sec. \ref{sec:overview}), MORFEO complements its standard temporal filters with an Adaptive Vibration Cancellation (AVC) framework. While this work focuses on the primary control loops, it is worth noting that the architecture natively supports the concurrent activation of several dozen independent AVC instances. 

Each instance can be dynamically tuned to target specific combinations of modes and frequencies, providing a highly modular solution for rejecting narrow-band structural resonances. This approach follows the well-established heritage of ESO's vibration \textcolor{black}{control strategies \cite{2012SPIE.8447E..12M, 2014ecc..conf..303M, 7320616, 2018SPIE10703E..1GO}}, which are widely and successfully utilized across various VLT instruments to maintain performance in the presence of complex mechanical disturbances.

\subsection{Integration Time Optimization}

The integration time ($\Delta t$) for the LO and \textcolor{black}{Ref} sensors is \textcolor{black}{optimized during the acquisition phase as a function of the NGS magnitude to maximize the signal-to-noise ratio. Once this optimal value is set and the loops are closed, the integration time remains strictly fixed for the entire duration of the observation.}
Tables \ref{tab:lo_int_time} and \ref{tab:r_int_time} summarize the baseline framerate values (in milliseconds) currently considered for the LO and \textcolor{black}{Ref} loops, respectively.
It is important to note that these values are approximate baselines and will be refined during the \textcolor{black}{Assembly, Integration, and Verification (AIV)} phase and that the selected integration times must always be integer multiples or sub-multiples of the HO loop cycle time \textcolor{black}{(nominal 2 ms)} to ensure synchronization.
Moreover, \textcolor{black}{for} bright NGSs (magnitudes brighter than those listed), a 1 ms integration time is available.

\begin{table}[ht]
\caption{Baseline integration times ($\Delta t$) for the Low Order WFSs as a function of H-band magnitude.}
\label{tab:lo_int_time}
\begin{center}
\begin{tabular}{|l|c|c|c|c|c|c|c|}
\hline
\rule[-1ex]{0pt}{3.5ex}  \textbf{H Magnitude} & \textcolor{black}{\textbf{$\leq$14}} & \textcolor{black}{\textbf{15}} & \textbf{16} & \textbf{17} & \textbf{18} & \textbf{19} & \textbf{20} \\
\hline
\rule[-1ex]{0pt}{3.5ex}  $\Delta t$ [ms] (full-aperture config.) & \textcolor{black}{1} & \textcolor{black}{1} & 2 & 2 & 4 & 4 & 6 \\
\hline
\rule[-1ex]{0pt}{3.5ex}  $\Delta t$ [ms] ($2 \times 2$ config.) & \textcolor{black}{1} & \textcolor{black}{2} &  2 & 4 & 4 & 6 & 8 \\
\hline
\end{tabular}
\end{center}
\end{table}

\begin{table}[ht]
\caption{Baseline integration times ($\Delta t$) for the Reference WFS (focus/truth sensing) as a function of R-band magnitude.}
\label{tab:r_int_time}
\begin{center}
\begin{tabular}{|l|c|c|c|c|c|}
\hline
\rule[-1ex]{0pt}{3.5ex}  \textbf{R Magnitude} & \textbf{$\leq$17} & \textbf{18} & \textbf{19} & \textbf{20} & \textbf{21} \\
\hline
\rule[-1ex]{0pt}{3.5ex}  $\Delta t$ [ms] & 10 & 12 & 24 & 50 & 72 \\
\hline
\end{tabular}
\end{center}
\end{table}

\subsection{Multi-rate Control and Partial Reconstruction}
\label{sec:multi_rate}

As described in the previous section, Sec.~\ref{sec:int_time}, the integration times for the three LO WFSs (and similarly for the three \textcolor{black}{Ref} WFSs) are optimized independently to maximize the signal-to-noise ratio.
Therefore, a critical aspect of the MORFEO control strategy is the ability to handle heterogeneous framerates among the Natural Guide Star sensors.
Therefore, the arrival of slope measurements at the Real-Time Computer (RTC) is asynchronous.

To manage this, the control loop operates at the maximum system frequency ($f_\mathrm{max}$, typically 1000 Hz for LO and 100 Hz for \textcolor{black}{Ref}).
At each control cycle $k$, the RTC evaluates the availability of fresh measurement vectors from each of the three sensors.
Based on a validity vector identifying the currently available measurements, the RTC selects a specific, pre-calculated reconstruction matrix $R_{sub}$ from a lookup table.

This approach implies that the modal reconstruction is dynamic and depends on the instantaneous availability of the data:
\begin{itemize}
    \item LO loop: the LO loop aims to control global tip-tilt, \textcolor{black}{as well as plate scale (magnification and anamorphism) and field rotation modes. Physically, these linear field distortions correspond to the second radial order (focus and astigmatism) located out of the pupil (at high altitudes). While these modes are captured geometrically by measuring the differential tip-tilt across the three LO WFSs, the control basis reconstructs them explicitly as a combination of high-altitude and ground-layer focus and astigmatism, designed to result in zero net focus and astigmatism along the lines of sight. The full observability of this set of modes is achieved only when measurements from all three LO WFSs are available simultaneously.}
    When only a subset of sensors provides data (e.g., during the integration time of a fainter star), a reduced reconstruction matrix is applied, estimating only the modes observable with the available geometry (e.g., global Tip-Tilt from a single sensor).   
    \item \textcolor{black}{Ref} loop: a similar logic applies to the Reference loop, which reconstructs approximately 50 low-order modes per layer to detrend LGS biases.
    The tomographic reconstruction of the full modal basis is performed only at the ``beat'' frequency where all three Reference WFSs have integrated their frames.
    In intermediate steps, the correction is updated using partial reconstruction matrices based on the faster reference stars, ensuring that the control loop remains stable and responsive even with faint sky coverage constellations.
\end{itemize}
%This strategy ensures that the photon flux from brighter stars is exploited at the highest possible rate to reject fast disturbances (like windshake on tip-tilt), while the information from fainter stars is seamlessly incorporated as soon as it becomes available to constrain the slower, field-dependent modes.
We acknowledge that switching between matrices implies a discontinuity in the error signal at the update frequency of the slower sensors, potentially inducing small oscillations on the common pupil modes. 
However, since the closed-loop bandwidth is a fraction of the frame rate, the loop transfer function acts as a low-pass filter, effectively smoothing these artifacts within the error budget.
Moreover, this multi-rate fusion is applied conservatively: it provides a genuine gain primarily when reference stars have comparable magnitudes. 
In cases of large frame rate disparities (e.g., ratio $>10$), the contribution of the slower, noisier sensor to the fast pupil modes is discarded, and high-order correction relies solely on the brighter stars.
Conversely, regarding the Tip-Tilt zero-point (DC component), measurements from all available sensors are averaged to define the correct pointing regardless of their frame rate.
This ensures the minimization of quasi-static optical relay distortions and field-dependent wobble over long observation windows, a task that does not require high-bandwidth correction.
A rigorous mathematical framework demonstrating how to stably fuse these asynchronous measurements using a Linear Time-Invariant (LTI) complementary filtering approach is detailed in Appendix~\ref{sec:multirate_lti}.
This dynamic matrix selection logic and asynchronous vector management have been fully implemented in the \texttt{SPECULA} framework\footnote{\url{https://github.com/ArcetriAdaptiveOptics/SPECULA}} through the \texttt{ModalRecMultirate} processing object, allowing seamless end-to-end simulations of MORFEO's heterogeneous WFS configurations.

%%%%%%%%%%%%%%%%%%%%%%%%%%%%%%%%%%%%%%%%%%%%%%%%%%%%%%%%%%%%%%%%%%%%%%%%%

\section{Command matrix}
\label{sec:comm_mat}

The command matrix is made of two parts following Fusco et al. 2001~\cite{Fusco2001}: a tomographic reconstruction and a projection operator.

The tomographic reconstruction is done on a set of layers distributed on the full range of atmospheric turbulence, typically from ground to about 20 km and it is provided by a minimum-mean-square-error (MMSE) estimator.
Given the turbulence covariance \(C_{\Phi}\), the noise covariance \(C_n\), and the interaction matrix \(D\), the estimator \(R\) is:
\begin{equation}\label{eq:mmse}
    R = \left( D^{\top} C_n^{-1} D + C_{\Phi}^{-1}\right)^{-1} D^{\top} C_n^{-1}
\end{equation}
For a linear model with Gaussian statistics, this MMSE estimator is equivalent to the Bayesian solution (the mean of the posterior distribution).
\textcolor{black}{The layer weights (effective $r_0$) populating the prior covariance matrix $C_\phi$ are determined by computing a baseline turbulence covariance from the von Kármán phase spectrum, which is then scaled for each fixed reconstruction altitude by the fractional $C_n^2$ weights derived from the official ESO atmospheric profile~\cite{2013aoel.confE..89S}.}
The calibration of the interaction matrices is performed considering a perfect knowledge of the geometry and the DM influence functions, but the WFS model used does not include the spot elongation.
The spot elongation model is included in the noise covariance \(C_n\) as shown in \textcolor{black}{B\'echet et al. 2010~\cite{2010JOSAA..27A...1B}}.
To efficiently generate these large-scale operators, we developed \texttt{SynIM}\cite{2026arXiv260607759A}, a dedicated Python package.
\texttt{SynIM} analytically computes the synthetic interaction matrices exploiting the exact system geometry, the theoretical WFS parameters, and the DM influence functions, drastically reducing the computational overhead compared to empirical push-pull methods.

The LGS WFSs are rotated relative to the pupil in order to minimize truncation and achieve super-resolution\cite{2022JATIS...8b1514F,2025ExA....60...26C}.
The parameters of the misalignments are estimated on-line by the system using the SPRINT algorithm\cite{2021MNRAS.504.4274H} as described in \textcolor{black}{Agapito et al. 2024~\cite{2024SPIE13097E..5PA}}. 
Please note that imperfect knowledge of the registration has an impact on the performance as shown in Agapito et al. 2020\cite{2020SPIE11448E..2SA}.

The projection operator finds the best DM commands to maximize the correction in a given field of view, as determined by a set of directions.
Its implementation is the same as that described by Fusco et al. 2001\cite{Fusco2001}, with the additional features of different weights on the directions and Tikhonov regularization in the pseudo-inverse. 
Typically, we use directions in the middle and at the edge of the science FoV, as well as some additional directions with low weights in the technical FoV, as can be seen in Table \ref{tab:proj}.
This is done to improve the image quality on the NGS WFSs, as the accuracy of the wavefront sensing—particularly the tip-tilt estimation in the LO WFS—is directly dependent on the spot sharpness and the resulting signal-to-noise ratio.

The final command matrix, $M_c$, used to compute the Deformable Mirror commands from the WFS slopes, is defined as the cascaded product of four specialized operators:
\begin{equation}
    \label{eq:comm_mat}
    M_c = \Pi_{\text{misreg}} \cdot P \cdot R \cdot F_{\text{TTF}} 
\end{equation}
where the sequence of operations from measurements to commands is structured as follows:
\begin{itemize}
    \item \textcolor{black}{$F_{\text{TTF}}$ is the \textbf{Tip-Tilt and Focus (TTF) filter matrix}. It mathematically projects out the global TT and Focus modes from the LGS slopes. This step is essential to ensure these modes are exclusively driven by the NGS loops, preventing the injection of sodium layer altitude biases and windshake aliasing from the LGS measurements.}
    \item $R$ is the \textbf{tomographic reconstruction matrix} (Eq.~\ref{eq:mmse}), which estimates the volumetric phase perturbations from the WFS measurements.
    \item $P$ is the \textbf{projection matrix}, mapping the reconstructed atmospheric layers onto ideal DM commands to maximize performance over the specified FoV.
    \item $\Pi_{\text{misreg}}$ is the \textbf{virtual-to-physical DM projection matrix}. It adapts the ideal DM commands to the actual physical actuator grid, actively compensating for dynamic mis-registrations (such as rotation or pupil shifts) identified by the online tracking algorithms.
\end{itemize}

Finally, note that the LO command matrix can cope with all possible NGS asterisms, including those with two or a single NGS.
The HO command matrix can also be calculated when one or two LGS are unavailable: in this case, the system will not provide the required performance, but it can still operate.

\begin{table}[ht]
\caption{Two examples of position and weight of the directions used to compute the Projection Operator. All directions are uniformly distributed on circles and weights are normalized to 1.}
\label{tab:proj}
\begin{center}
\begin{tabular}{|c|c|c|c|}
\hline
\rule[-1ex]{0pt}{3.5ex}  \multirow{2}{*}{\textbf{Configuration}} & \textbf{Circle Radius} & \textbf{\textcolor{black}{Number of} Directions} & \textbf{Weight per} \\
\rule[-1ex]{0pt}{3.5ex}  {} & \textbf{[arcsec]} & \textbf{on the Circle} & \textbf{Direction} \\
\hline
\rule[-1ex]{0pt}{3.5ex}  \multirow{3}{*}{large science FoV} & 0 & 1 & 0.0562 \\
\cline{2-4}
\rule[-1ex]{0pt}{3.5ex}  {} & 30 & 8 & 0.1124 \\
\cline{2-4}
\rule[-1ex]{0pt}{3.5ex}  {} & 80 & 8 & 0.0056 \\
\hline
\rule[-1ex]{0pt}{3.5ex}  \multirow{3}{*}{small science FoV} & 0 & 1 & 0.0562 \\
\cline{2-4}
\rule[-1ex]{0pt}{3.5ex}  {} & 10 & 8 & 0.1124 \\
\cline{2-4}
\rule[-1ex]{0pt}{3.5ex}  {} & 80 & 8 & 0.0056 \\
\hline
\end{tabular}
\end{center}
\end{table}

\subsection{WFS Pupil Image Control}
\label{sec:pup_control}

\textcolor{black}{In MORFEO, pupil stabilization is strictly required for the high-order LGS WFSs and the intermediate-order Ref WFSs. To achieve this, each of the 9 WFS opto-mechanical paths is equipped with a mechanism to optically recenter the pupil image. For the Natural Guide Star channels, the LO and Ref WFSs share a common optical path; therefore, the pupil is inherently stabilized for the LO sensors as well, despite their configurations ($2 \times 2$ or LIFT) being largely insensitive to small pupil shifts. For the LGS and Ref sensors, this functionality ensures that the definition of valid sub-apertures (i.e., the specific set of detector pixels used to compute the wavefront gradient) remains constant during operations and that each sub-aperture receives proper illumination.}

It is crucial to distinguish this optical centering from the maintenance of a perfect alignment between the Deformable Mirror (DM) actuators and the WFS sub-apertures.

\begin{figure}[h!]
    \centering
    \includegraphics[width=0.6\linewidth]{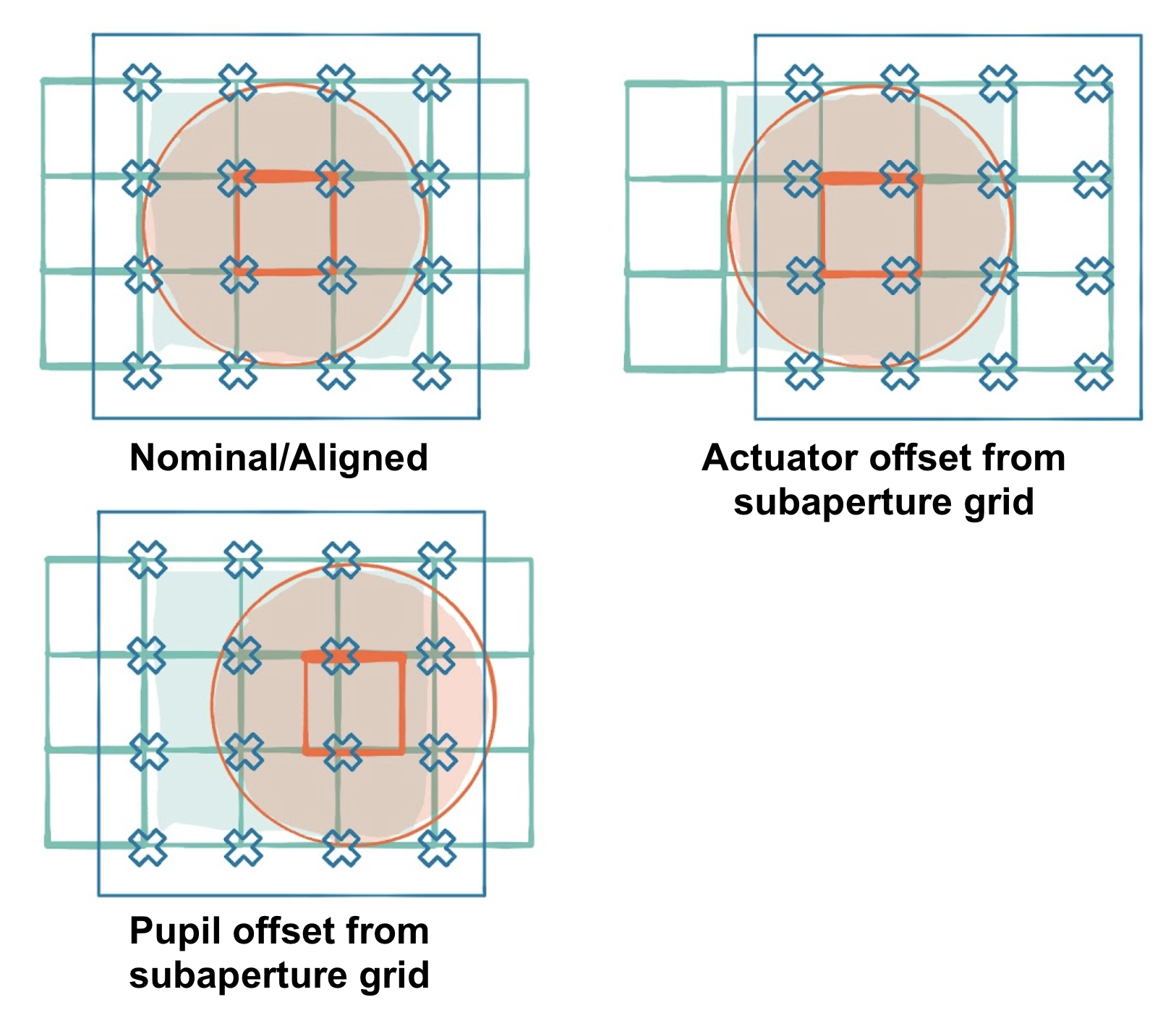}
    \caption{
    Conceptual distinction between pupil centering and actuator-subaperture matching. 
    \textcolor{black}{The blue crosses represent the Deformable Mirror (DM) actuators. \textbf{Top-left} ("Nominal/Aligned"): nominal alignment with the pupil footprint centered over the lenslet array ($3\times3$ grid) and DM actuators ($4\times4$ grid) aligned in a Fried geometry. \textbf{Top-right} ("Actuator offset from subaperture grid"): the pupil centering is maintained, but the relative alignment between actuators and sub-apertures is lost. In MORFEO, this affects \textbf{all DMs} and originates from hardware tolerances (e.g., the $\sim 1\%$ wandering of M4 actuators) and the complex system kinematics: post-focal DMs rotate relative to the LGS WFSs (which track M4), the LGS rotator introduces run-out, and the sky-tracking NGS WFSs rotate with respect to the telescope and post-focal relay. \textbf{Bottom-left} ("Pupil offset from subaperture grid"): the DM actuators still match the sub-aperture grid, but the pupil footprint is decentered. This effect is primarily driven by telescope optics, gravity flexures, and active tracking movements, causing the overall footprint to wander across the wavefront sensors.}    
    }
    \label{fig:pup_control}
\end{figure}

In the MORFEO system, all conditions illustrated in Fig.~\ref{fig:pup_control} can occur.
While the ELT requirements specify a pupil lateral stability of approximately 0.8\% (roughly half a sub-aperture), the actual position is subject to drifts during an observation due to mechanical flexures and telescope collimation adjustments.
Consequently, there is no guarantee that the pupil footprint on M4 remains stable relative to the actuator grid to within a small fraction of the actuator pitch.

A downstream instrument cannot mechanically decouple these effects: optically recentering the pupil footprint on the lenslet array inherently shifts the projection of the DM actuators relative to the sub-apertures.
Thus, the adopted control strategy separates the compensation mechanisms:
\begin{itemize}
    \item \textbf{Hardware Correction:} each MORFEO WFS is equipped with physical actuators (pupil steering mirrors) to actively recenter the pupil footprint on the lenslet array, ensuring constant illumination.
    \item \textbf{Numerical Correction:} the mismatch between actuators and sub-apertures is handled numerically. When the misalignment exceeds a critical threshold, the control matrices (interaction and reconstruction matrices) are recomputed to account for the new geometry (using SPRINT as described in Agapito et al. 2024\cite{2024SPIE13097E..5PA}).
\end{itemize}

The measurement of the pupil footprint position on the WFS is based on pupil shape fitting and it is described in  Lampitelli et al. 2024\cite{2024SPIE13101E..3TL}.

\subsection{Reconstruction Reference System and Matrix Updates}
\label{sec:update}

\textcolor{black}{The MORFEO control strategy must manage a complex and variable geometry} between the telescope pupil, the guide stars, and the deformable mirrors.

For the High Order (HO) loop, the wavefront is reconstructed on layers at fixed \textcolor{black}{distances along the telescope line of sight. This fixed-geometry architecture is a deliberate computational optimization: tracking the true atmospheric altitudes as the zenith angle varies would dynamically alter the meta-pupil size, necessitating multiple sets of virtual mirror shapes and the continuous recomputation of the projection matrix $P$ (see Eq.~\ref{eq:comm_mat}). A fixed reference frame allows for a single $P$ matrix and streamlines the interaction matrix calibration.}
Thanks to the LGS WFS de-rotator, the telescope pupil, laser beacons, and M4 remain \textcolor{black}{rotationally} static relative to each other.
However, two elevation-dependent effects require periodic updates of the control matrices:
\begin{itemize}
    \item The rotation of the Post-Focal Deformable Mirrors (PFDMs) relative to the WFS sub-apertures.
    \item The expansion of the LGS footprint on high-altitude layers as the airmass increases.
\end{itemize}
Geometric analysis of the footprint expansion rate shows that to keep the mismatch below $1/10^{th}$ of the actuator pitch \textcolor{black}{(strictly limiting the error to $<10$ nm RMS), the HO reconstruction matrix must be updated approximately every 6 minutes for observations at a $30^\circ$ zenith angle. In the most demanding geometric regime ($60^\circ$ zenith angle), maintaining this $<10$ nm threshold requires an update approximately every 4 minutes. It is worth noting that this represents a conservative operational limit; at higher zenith angles, the increased airmass naturally relaxes the absolute error budget, potentially allowing for slightly longer update intervals without significant relative performance degradation.}

Contrary to the LGSs, the Natural Guide Stars (NGSs) rotate with the field. This necessitates a more dynamic update strategy:
\begin{itemize}
    \item Low Order Loop: the critical constraint here is imposed by the LO tip-tilt control. The system must correct windshake disturbances with amplitudes up to $50\,\mu m/s$.
    During field rotation a projection error will be made between tip and tilt.
    While the closed loop effectively rejects most of this projection error, the temporal error leaves an uncorrected residual.
    To maintain the performance budget, we require that the additional error introduced by the projection misalignment (cross-coupling between tip and tilt axes) remains below $10\,\text{nm}$ RMS. This implies a maximum tolerable projection error of $2\%$ ($10/500$), which corresponds to a maximum rotation angle of approximately $1.15^\circ$ and an update time of 8s, driven by the maximum field rotation rate encountered when pointing close to the zenith (the worst-case scenario).
    \item Reference Loop: the Reference loop is less sensitive to fast rotation due to its lower bandwidth, the nature of the higher-order corrections involved, and the larger sub-aperture size of the \textcolor{black}{Ref} WFS. An update rate of the reconstructor every 30 seconds is sufficient to keep the rotation error negligible (below $1/10^{th}$ of a sub-aperture).
\end{itemize}
This multi-scale update strategy optimizes the computational load on the Real-Time Computer, reserving high-frequency updates only for the small, critical matrices of the LO loop.

%%%%%%%%%%%%%%%%%%%%%%%%%%%%%%%%%%%%%%%%%%%%%%%%%%%%%%%%%%%%%%%%%%%%%%%%%

\section{Conclusion}
\label{sec:conclusion}

The control strategy for MORFEO described in this work represents the consolidated baseline for the Final Design Review.
The architecture has been tailored to tackle the specific challenges of the ELT, transitioning from preliminary trade-offs to a robust, implementation-ready design.

Major architectural definitions include the Focus Control Strategy, which now relies entirely on Natural Guide Stars to decouple the system from the variability of the Sodium layer altitude.
The adoption of a hybrid Low-Order WFS configuration (combining Full-Aperture flux sensitivity with 2$\times$2 SHS focus capability) proves to be the optimal solution to maximize Sky Coverage while maintaining robust tip-tilt and focus correction.
Additionally, the strategy successfully integrates advanced features such as petaling control via LIFT and optimized LGS spot truncation mitigation, ensuring that the system performance is not limited by the specific geometry of the ELT pupil or the LGS properties.

End-to-end simulations (performed with PASSATA\cite{2016SPIE.9909E..7EA} and more recently with SPECULA\cite{specula2026}) confirm that this control architecture allows MORFEO to meet its rigorous performance requirements in terms of Strehl Ratio and astrometric precision across the technical field.
The project is now entering the construction phase, where these algorithms will be implemented and tested on the final RTC hardware.

%%%%%%%%%%%%%%%%%%%%%%%%%%%%%%%%%%%%%%%%%%%%%%%%%%%%%%%%%%%%%%%%%%%%%%%%%

\subsection* {Acknowledgments}
The authors wish to thank all the members of the MORFEO consortium for their continuous support and fruitful discussions.
Finally, the authors acknowledge the use of Gemini (Google) for English language refinement and formatting assistance during the preparation of this manuscript.

\section* {Disclosures}
The authors declare there are no financial interests, commercial affiliations, or other potential conflicts of interest that have influenced the objectivity of this research or the writing of this paper.

\subsection* {Materials and Methods}
The performance estimation and validation of the control strategies described in this work were conducted through high-fidelity end-to-end numerical simulations using the PASSATA and SPECULA  frameworks. The analytical derivations of the control filters were performed using standard Z-domain discrete-time analysis.

\subsection* {Code and Data Availability} 
The data and the code underlying this article will be shared upon reasonable request to the corresponding author (guido.agapito@inaf.it), subject to the approval of the relevant scientific collaborations.

\newpage

\appendix

\section{IIR filter for Tip-Tilt}
\label{sec:iir_tt_app}

\begin{figure}[!h]
    \centering
    \includegraphics[width=0.9\linewidth]{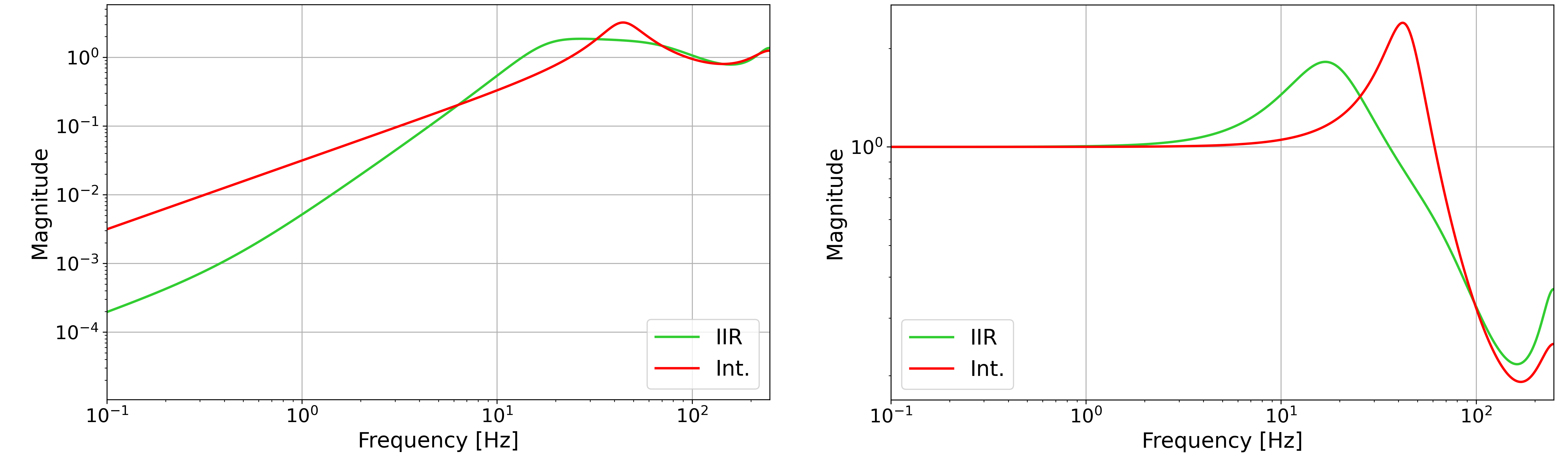}
    \caption{Closed loop Transfer Functions (TFs) considering a plant that is a delay of three frames. Left, Rejection TF, right, Noise TF. Comparison of an integrator control and an IIR control with two poles (1.000 and 0.995) and two zeros (0.450 and 0.850). The gain is 0.4 for both controllers.}
    \label{fig:RTFandNTF}
\end{figure}

The filter is implemented in the z-domain and is defined by an effective yet simple structure of two poles and two zeros.
\textcolor{black}{This structure serves as a second-order integrator-lead compensator, where $g$ represents the overall gain of the filter:}
\begin{equation}
   H(z) = g \cdot \frac{(z - z_1)(z - z_2)}{(z - p_1)(z - p_2)} 
\end{equation}
The selected coefficients\footnote{note that these are the same coefficients used in the LO part of TipTop\cite{2020SPIE11448E..2TN}.} are: $p_1 = 1.000$, $p_2 = 0.995$, $z_1 = 0.850$ and $z_2 = 0.450$,
\textcolor{black}{and its Closed Loop Transfer Functions are shown in Fig.~\ref{fig:RTFandNTF}.}
The pole placement is critical for achieving high rejection while maintaining stability.
We placed one pole exactly at $z = 1.000$ (on the unit circle, $p_1 = 1.000$) which introduces an ideal integrator into the control loop. This ensures \textbf{infinite steady-state gain} at DC (0 Hz), guaranteeing aggressive rejection for the dominant low-frequency wind shake component.
The second pole is placed slightly inside the unit circle, at $p_2 = 0.995$.
This slight offset is necessary to prevent the closed-loop resonance peak from shifting excessively close to DC.
\textcolor{black}{This is a critical stability consideration, especially since the plant dynamics are approximated with a worst-case time delay of 3 frames (occurring at the maximum loop frequency of 1 kHz; at a standard 500 Hz frame rate, the delay is approximately 2 frames), which severely limits the stable gain margin for frequencies close to zero.}
The two zeros are primarily used to shape the loop's response at higher frequencies, providing phase lead and controlled roll-off: the zeros limit the propagation of high-frequency measurement noise from the WFSs into the DM commands.
They introduce a damping effect, ensuring a safe loop gain roll-off after the critical crossover frequency.
These zeros are instrumental in precisely shaping the closed-loop bandwidth and attenuating specific high-frequency components or residual oscillations.
Beyond the filter structure, we are also actively investigating advanced optimization techniques, specifically focusing on \textbf{gain optimization} strategies \cite{2021MNRAS.508.1745A,FaloticoAO4ELT8} and full control law optimization to prepare for the final on-sky calibration and deployment at the telescope.

%%%%%%%%%%%%%%
%%%%%%%%%%%%%%
%%%%%%%%%%%%%%
%%%%%%%%%%%%%%
%%%%%%%%%%%%%%

\section{Linear Time-Invariant Approximation for Multirate Sensor Fusion}
\label{sec:multirate_lti}

To rigorously analyze the stability and the dynamics of fusing measurements from Natural Guide Stars operating at different framerates, we can model the system using a complementary filtering approach. Let us consider the generalized MORFEO configuration featuring one fast sensor $y_f$ (representing the brightest NGS) operating at the fast loop period $T_f$, and a set of $N_\mathrm{slow}$ fainter sensors $y_{si}$, each operating at a slower period $T_{s,i} = N_i \cdot T_f$.

WFSs directly measure the tracking error.
Assuming the sensors exhibit static biases given by NCPAs or atmospheric anisoplanatism, an estimation of the physical tracking error $e$ can be provided by the barycentric average of the available sensors:
\begin{equation}
    e \approx W_f y_f + \sum_{i=1}^{N_\mathrm{slow}} W_{si} y_{si}
\end{equation}
where $W_f$ and $W_{si}$ are normalized DC weights (e.g., $1/3$ for each sensor in a uniform 3-NGS triplet). 

\subsection{Dynamic Offset Estimation and Zero-Stuffing}
To fuse these asynchronous measurements without injecting Zero-Order Hold (ZOH) transient steps into the fast primary loop, the slow framerate is used strictly to compute a quasi-static bias correction (offset). At the slow rate $T_{s,i}$, the discrepancy is processed through a slow integral controller with gain $g_s$. 

To express this multi-rate logic entirely in the fast discrete-time $Z$-domain, we must mathematically define the downsampling and upsampling process. This is achieved via the zero-stuffing operator $\mathcal{S}_{N_i}$, which maps the slow sequence into the fast-rate domain by inserting $N_i-1$ zeros between each valid sample.
Applying the Z-transform, this complex modulation is:
\begin{equation}
    \mathcal{S}_{N_i} \{ y_{si}(z) \} = \frac{1}{N_i} \sum_{k=0}^{N_i-1} y_{si} \left( z e^{-j \frac{2\pi k}{N_i}} \right)
\end{equation}
The zero-stuffed signal is then acted upon by the ZOH operator $H_{N_i}(z) = \frac{1 - z^{-N_i}}{1 - z^{-1}}$ to construct the physical piecewise-constant staircase signal fed to the control loop.

\subsection{LTI Control Law and Natural Stability}
By substituting the ZOH and the integral controllers into the closed-loop equations, a fundamental pole-zero cancellation occurs: the comb-like numerator of the ZOH operator ($1 - z^{-N_i}$) perfectly cancels the resonant pole of the slow integrator. 

To understand this transition physically, we can view the zero-stuffing operator $\mathcal{S}_{N_i}$ as converting the slow-rate WFS measurements into a sparse, ``comb-like'' train of impulses at the fast clock rate. While generating this sparse comb mathematically introduces high-frequency spectral replicas (aliasing images), the physical integration of the slow WFS over its extended exposure time inherently acts as a moving-average anti-aliasing filter. This natural low-pass filtering heavily suppresses the high-frequency artifacts. 

By discarding these attenuated replicas, the complex, multi-branch (``forked'') multi-rate architecture simplifies into a pure baseband Linear Time-Invariant (LTI) equivalent model. Under this approximation, the parallel control paths merge, and the total cascaded control effort $U(z)$ collapses elegantly into a shared double-integrator formulation:
\begin{equation}
    U(z) = \frac{C_f(z)}{1 - z^{-1}} \left( \left[ \left( 1 + g_\mathrm{track} W_f \right) - z^{-1} \right] y_f(z) + \sum_{i=1}^{N_\mathrm{slow}} \left[ g_\mathrm{track} W_{si} N_i \right] y_{si,\mathrm{stuffed}}(z) \right)
\end{equation}
where $C_f(z)$ is the primary fast controller (e.g., $g_f / (1-z^{-1})$), $g_\mathrm{track}$ is the global offset tracking gain, and $y_{si,\mathrm{stuffed}}$ are the mathematically sparse, zero-stuffed slow measurements.

This factorization reveals the structural stability of the complementary multirate approach:
\begin{itemize}
    \item Regardless of the number of asynchronous WFSs, the overall system collapses into a single shared integrator core $\frac{C_f(z)}{1 - z^{-1}}$.
    \item Each slow sensor contributes purely as a static, zero-stuffed proportional injection.
    \item The fast sensor $y_f$ is processed by a first-order FIR lead filter. This effectively generates a minimum-phase zero strictly bounded within the unit circle ($z_\mathrm{zero} = 1 / (1 + g_\mathrm{track} W_f) < 1$). 
\end{itemize}

\begin{figure}[ht]
    \centering
    \includegraphics[width=0.6\linewidth]{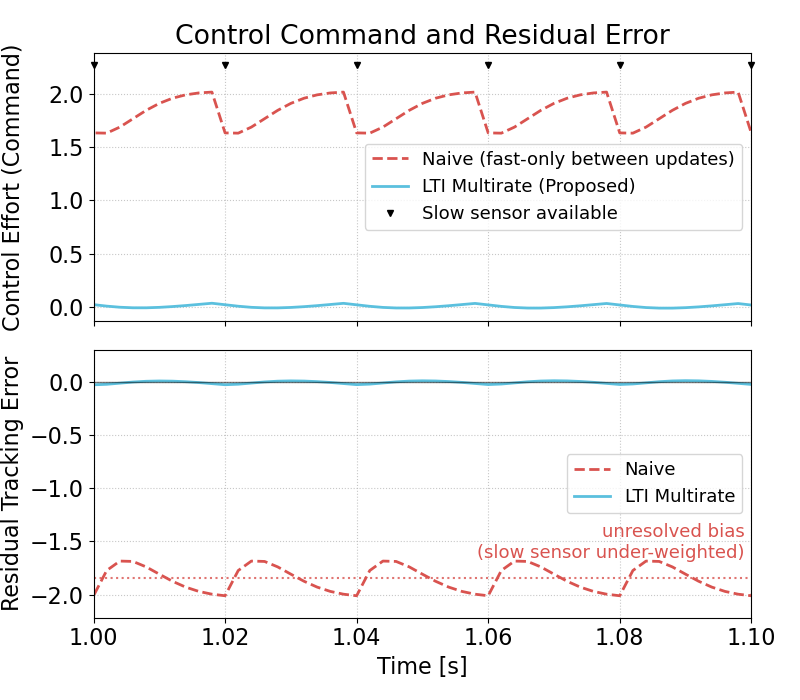}
    \caption{\textcolor{black}{Temporal simulation of the control behavior comparing a naive asynchronous update approach (red dashed line) and the proposed LTI multirate fusion (solid blue line) in the presence of uncalibrated static biases. Top: The naive approach (which processes the slow sensor only at its arrival and relies solely on the fast sensor between updates) injects "kick-and-drift" artifacts into the actuator commands, whereas the LTI formulation guarantees a smooth control effort. Bottom: The LTI multirate architecture smoothly drives the residual differential tracking error to zero. In contrast, the naive approach severly under-weights the slow sensor, leaving a significant unresolved steady-state bias alongside high-frequency harmonic ripples.}}
    \label{fig:LTI_multirate}
\end{figure}

This formulation mathematically guarantees that fusing multiple NGSs at disparate framerates remains structurally robust, automatically rejecting steady-state sensor biases without requiring complex dynamic retuning of the real-time computer matrices. \textcolor{black}{In addition, the shared integration naturally drives the steady-state differential tracking error to zero. Unlike naive asynchronous update approaches (e.g., simply dropping the slow measurements between updates) that severely under-weight the slow sensors and inject ``kick-and-drift'' harmonic ripples into the actuator commands, this LTI formulation provides smooth temporal tracking, minimizing the residual dynamic wavefront error (as demonstrated in Fig.~\ref{fig:LTI_multirate}).}

The practical realization of this complementary filtering architecture is available in the \texttt{SPECULA} framework as the \texttt{MultirateComplementaryFilter} class. This module automatically handles the zero-stuffing operations, the ZOH dynamics, and the cascaded integrations for an arbitrary number of asynchronous sensors.

%%%%%%%%%%%%%%
%%%%%%%%%%%%%%
%%%%%%%%%%%%%%
%%%%%%%%%%%%%%
%%%%%%%%%%%%%%

\section{The Steady-State in a Pseudo-Open Loop Control Approach}
\label{sec:polc_ss}

In this Appendix, we analyse the effect of the pseudo-open loop (POL) approach on closed-loop control, building on what has been presented in Busoni et al. 2019\cite{Busoni2019}.
The dynamics of an adaptive optics control system with a generic temporal filter \(\Gamma(z)\) can be described in the Z-domain. The relationship between the sensor measurements \(s_{meas}\) and the commands \(c\) is given by:
\begin{equation}\label{eq:controller_generale}
    c(z) = \left(I + \Gamma(z) H z^{-1}\right)^{-1} \Gamma(z) P R~s_{meas}(z)
\end{equation}
where \(H = I - P R D_{DM}\) is the second POL matrix and it represents the error propagation ($I$ is the identity matrix, \textcolor{black}{$R$ and $P$ are the reconstruction and projection matrices defined in Eq.~\ref{eq:mmse} and Sec. ~\ref{sec:comm_mat} respectively} and $D_{DM}$ is the DMs interaction matrix).

\subsection{Steady-State Behavior}

To analyze the steady-state behavior (DC gain), we set \(z \to 1\). The result depends on the nature of the filter \(\Gamma\). If the controller has an integrator (a pole in zero), its transfer function at \(z=1\) is infinite. This infinite gain is necessary to null the static error. In this case, the limit of the controller's transfer function becomes:
\begin{equation}
    \lim_{z\to1} \left(I+ \Gamma(z) Hz^{-1}\right)^{-1}\Gamma(z) = H^{-1}
\end{equation}
Consequently, the overall static gain of the system \(T_{ss}\) from the input \(s_{meas}\) to the output \(c\) is:
\begin{equation}\label{eq:T_ss}
    T_{ss} = \frac{c}{s_{meas}} \bigg|_{z=1} = H^{-1} P R = \left(I - P R D_{DM}\right)^{-1} P R
\end{equation}

\subsection{Simplification with a Perfect Model}

A fundamental simplification emerges when we assume that the reconstructed layers corresponds to the DMs, i.e., $D_{DM} = D$ and $P=I$ (there is no modal projection). Under these conditions, we can analyze the term $(I - R D)$ from Eq.~\eqref{eq:T_ss}.

The product \(RD\) simplifies significantly:
\begin{align}
    R D &= \left( D^{\top} C_n^{-1} D + C_{\Phi}^{-1}\right)^{-1} \left( D^{\top} C_n^{-1} D \right) \nonumber \\
    &= I - \left( D^{\top} C_n^{-1} D + C_{\Phi}^{-1}\right)^{-1} C_{\Phi}^{-1} \label{eq:RD_simplified}
\end{align}
where the matrix identity \((A+B)^{-1}A = I - (A+B)^{-1}B\) has been used. Substituting this result into \((I - R D)\):
\begin{equation}
    I - R D = \left( D^{\top} C_n^{-1} D + C_{\Phi}^{-1}\right)^{-1} C_{\Phi}^{-1}
\end{equation}
The inverse of this term, which appears in Eq.~\eqref{eq:T_ss}, is therefore:
\begin{equation}
    \left(I - R D\right)^{-1} = C_{\Phi} \left( D^{\top} C_n^{-1} D + C_{\Phi}^{-1}\right)
\end{equation}
Finally, assembling everything to calculate the steady-state gain \(T_{ss} = (I - RD)^{-1} R\):
\begin{align}
    T_{ss} &= \left[ C_{\Phi} \left( D^{\top} C_n^{-1} D + C_{\Phi}^{-1}\right) \right] \left[ \left( D^{\top} C_n^{-1} D + C_{\Phi}^{-1}\right)^{-1} D^{\top} C_n^{-1} \right] \nonumber \\
    &= C_{\Phi} D^{\top} C_n^{-1} \label{eq:final_result}
\end{align}
The final result is open to a physical interpretation.
The term \( C_{\Phi} D^{\top} C_n^{-1} \) represents the projection of the noise-weighted measurements (\( D^{\top} C_n^{-1} \)) onto the turbulence subspace defined by the covariance \( C_{\Phi} \).
\textcolor{black}{Essentially, this term acts as an SNR-weighted estimator.}
It indicates that, at steady state, the closed-loop control effort is proportional to the expected turbulence power (\( C_{\Phi} \)) and inversely proportional to the measurement noise (\( C_n \)).
Thus, for modes with high SNR, the system applies a strong correction, whereas for modes dominated by noise (or with low turbulence energy), the steady-state gain is naturally suppressed, effectively acting as a spatial regularization that complements the temporal filtering.

This behavior is mathematically analogous to a multi-dimensional leaky integrator.
While a standard integrator possesses infinite DC gain to perfectly reject static disturbances, the MMSE-based control introduces a ``leakage'' term governed by the SNR.
For a given mode $i$, the effective steady-state gain approximates the Wiener filter ratio:
\begin{equation}
    G_{i, ss} \approx \frac{\sigma_{\text{turb}, i}^2}{\sigma_{\text{turb}, i}^2 + \sigma_{\text{noise}, i}^2}
\end{equation}
where $\sigma_{\text{turb}}^2$ and $\sigma_{\text{noise}}^2$ are the variances of the turbulence and noise for that mode, derived from the diagonals of $C_{\Phi}$ and $C_n$. Consequently, the system naturally relaxes the correction for noisy modes, preventing noise amplification at the cost of a finite residual static error, which is the optimal strategy in a minimum-variance sense.

\textcolor{black}{This emergent leaky integrator behavior is of extreme practical relevance for the stabilization of specifically ill-posed or unobservable modes in next-generation segmented telescopes. For instance, a conceptually equivalent temporal regularization (the MMSE Soft-Limiter) has been recently demonstrated by Agapito et al.\cite{Agapito2026_submitted} to successfully mitigate the "Island Effect" in SCAO systems. By applying a selective forgetting factor governed by von K\'arm\'an statistical priors strictly to the differential piston modes, the loop effectively acts as a localized leaky integrator. This prevents the control-induced divergence of poorly sensed petal modes without penalizing the correction of continuous atmospheric turbulence, confirming the robustness of regularized pseudo-open loop architectures.}

\subsection{Generalization to Tomographic Projection}

In the general case where the simplifications $D_{DM} = D$ and $P=I$ do not hold (e.g., when the number of reconstructed layers differs from the number of DMs, or when optimizing for a specific Field of View), the exact algebraic cancellation of Eq.~\eqref{eq:final_result} is not possible. However, the physical interpretation of the control behavior remains consistent.

Starting from the general steady-state gain in Eq.~\eqref{eq:T_ss}:
\begin{equation}
    T_{ss} = \left(I - P R D_{DM}\right)^{-1} P R
\end{equation}
We can interpret the role of the reconstructor \(R\) (the MMSE estimator) separately from the projection \(P\). The MMSE estimator \(R\) inherently produces a regularized estimate of the turbulent phase volume, \(\hat{\phi}_{vol}\), where noisy modes are suppressed (the ``leakage'' effect described above).
\begin{equation}
    \hat{\phi}_{vol} = R s_{meas} \approx \underbrace{C_{\Phi} D^{\top} \left(D C_{\Phi} D^{\top} + C_n\right)^{-1}}_{\text{Wiener Filter}} s_{meas}
\end{equation}
The term \((I - P R D_{DM})^{-1} P\) effectively acts as a fitting operator that maps this regularized volumetric estimate onto the available Deformable Mirrors.
Assuming the system is stable and the projection \(P\) is optimized to minimize the fitting error in the scientific FoV, the steady-state command \(c_{ss}\) tends towards the best fit of the MMSE estimate:
\begin{equation}
    c_{ss} \approx \text{Proj}_{DM} \left( \hat{\phi}_{MMSE} \right)
\end{equation}
Therefore, even in the general case, the ``leaky integrator'' behavior is preserved: the regularization embedded in \(R\) ensures that the volumetric estimate \(\hat{\phi}\) is conservative in low-SNR regimes. The projection \(P\) then transmits this conservative approach to the DMs. Therefore, the control loop will still naturally reject noise amplification for poorly sensed modes, effectively realizing a Projected Wiener Filter.
This Pseudo-Open Loop Control architecture, including the explicit computation of the $H$ error propagation matrix and the temporal update of the steady-state equations, has been natively implemented in the \texttt{SPECULA} simulation suite via the \texttt{ModalRecImplicitPolc} processing object.

%%%%% References %%%%%

\bibliography{report}   % bibliography data in report.bib
\bibliographystyle{spiejour}   % makes bibtex use spiejour.bst

%%%%% Biographies of authors %%%%%

\vspace{2ex}\noindent\textbf{Guido Agapito} Guido Agapito is a senior researcher at INAF, Arcetri Astrophysical Observatory, with more than 15 years of experience in AO for astronomy. He was involved in the design and commissioning of AO instruments such as VLT ERIS and LBT SOUL and he is currently involved in ELT MORFEO and VLT MAVIS. He is an expert in AO control systems, the main developer of the PASSATA and a developer of the analytical AO simulation framework TipTop.

\vspace{1ex}
\noindent Biographies and photographs of the other authors are not available.

%\listoffigures
%\listoftables

\end{spacing}
\end{document}